\documentclass[sigconf]{acmart}
\AtBeginDocument{%
  }

\copyrightyear{2026}
\acmYear{2026}
\setcopyright{cc}
\setcctype{by}
\acmConference[KDD '26]{Proceedings of the 32nd ACM SIGKDD Conference on Knowledge Discovery and Data Mining V.2}{August 09--13, 2026}{Jeju Island, Republic of Korea}
\acmBooktitle{Proceedings of the 32nd ACM SIGKDD Conference on Knowledge Discovery and Data Mining V.2 (KDD '26), August 09--13, 2026, Jeju Island, Republic of Korea}
\acmDOI{10.1145/3770855.3818964}
\acmISBN{979-8-4007-2259-2/2026/08}

\usepackage{makecell} 
\usepackage{caption}

\usepackage{algorithm}      
\usepackage{algorithmic}    

\usepackage{booktabs}
\usepackage{multirow}
\usepackage{colortbl}
\usepackage{enumitem}
\usepackage{xcolor}
\usepackage{pdfpages}
\usepackage{amsmath}
\usepackage{graphicx}
\usepackage{gensymb}
\usepackage{subcaption}

\definecolor{background_gray}{gray}{0.9}

\usepackage{amsthm}             

\usepackage{newfloat}
\usepackage{listings}

\begin{document}

\title{Hierarchical Reinforcement Learning for Cooperative Air-Ground Delivery in Urban System
}


\author{Songxin Lei}
\affiliation{%
  \institution{The Hong Kong University of Science and Technology (Guangzhou)}
  \city{Guangzhou}
  \country{China}}
\email{slei924@connect.hkust-gz.edu.cn}

\author{Chunming Ma}
\affiliation{%
  \institution{The Hong Kong University of Science and Technology (Guangzhou)}
  \city{Guangzhou}
  \country{China}}
\email{cma859@connect.hkust-gz.edu.cn}

\author{Haomin Wen}
\affiliation{%
  \institution{Carnegie Mellon University}
  \city{Pittsburgh}
  \state{Pennsylvania}
  \country{USA}}
\email{wenhaomin.whm@gmail.com}

\author{Yexin Li}
\affiliation{%
  \institution{Beijing Institute for General Artificial Intelligence}
  \state{Beijing}
  \country{China}}
\email{yliby@connect.ust.hk}

\author{Lizhenghe Chen}
\affiliation{%
  \institution{The Hong Kong University of Science and Technology (Guangzhou)}
  \city{Guangzhou}
  \country{China}}
\email{lizhenghec@hkust-gz.edu.cn}

\author{Qianyu Yang}
\affiliation{%
  \institution{Beijing Institute of Technology}
  \state{Beijing}
  \country{China}}
\email{yangqy@bit.edu.cn}

\author{Fugee Tsung}
\affiliation{%
  \institution{The Hong Kong University of Science and Technology}
  \state{Hong Kong SAR}
  \country{China}}
\email{season@ust.hk}

\author{Lei Chen}
\affiliation{%
  \institution{The Hong Kong University of Science and Technology (Guangzhou)}
  \city{Guangzhou}
  \country{China}}
\email{leichen@cse.ust.hk}

\author{Sijie Ruan}
\authornote{Corresponding authors.}
\affiliation{%
  \institution{Beijing Institute of Technology}
  \state{Beijing}
  \country{China}}
\email{sjruan@bit.edu.cn}

\author{Yuxuan Liang}
\authornotemark[1]
\affiliation{%
  \institution{The Hong Kong University of Science and Technology (Guangzhou)}
  \city{Guangzhou}
  \country{China}}
\email{yuxliang@outlook.com}

\renewcommand{\shortauthors}{Songxin Lei et al.}

\begin{abstract}

Cooperative air-ground delivery has emerged as a promising logistics paradigm by leveraging the complementary strengths of UAVs and ground carriers. However, effective dispatching in such heterogeneous systems faces two critical challenges: $i)$ the heterogeneity between flight and road dynamics, $ii)$ the scalability bottleneck raised by the exponential decision variables in large-scale fleets. To address these challenges, we propose HRL4AG, a \underline{H}ierarchical \underline{R}einforcement \underline{L}earning framework for cooperative \underline{A}ir-\underline{G}round delivery. 
Specifically, HRL4AG employs a high-level manager to tackle the scalability bottleneck by decomposing the joint action space, and mode-specific workers that encode distinct flight and road dynamics to address the heterogeneity.
Furthermore, a novel internal reward mechanism is designed to guide the hierarchical policy learning, addressing the credit assignment problem in sparse-reward settings. Extensive experiments on two real-world datasets and an evaluation platform demonstrate that HRL4AG significantly outperforms state-of-the-art baselines, improving the delivery success rate by up to 26\% while achieving an 80-fold increase in computational efficiency. Our source codes are available at \url{https://github.com/thunderlrr/Cooperative-Air-Ground-Delivery}.

\end{abstract}





\maketitle

\section{Introduction}

\begin{figure}[!t] 
    \centering
    \includegraphics[width=0.46\textwidth]{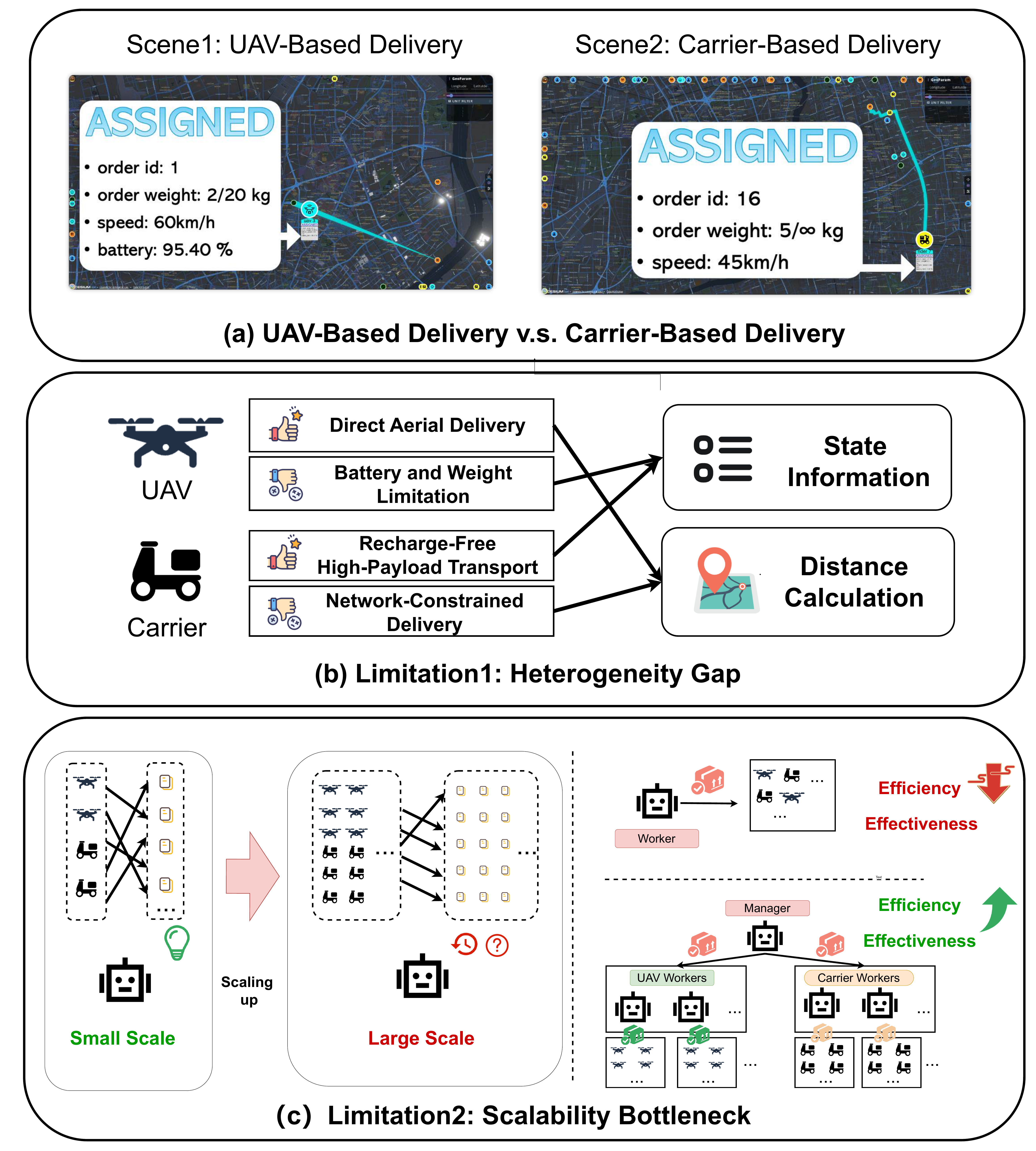}
    \caption{\textbf{Characteristics and Limitations of Cooperative Air-Ground Delivery. (a) Complementary nature of heterogeneous transportation modes. (b) Limitation 1: Heterogeneity gap. (c) Limitation 2: Scalability bottleneck.}}
    \label{fig:intro}
    \vspace{-2em}
\end{figure}

The rapid proliferation of on-demand logistics has catalyzed the exploration of novel delivery paradigms to satisfy the growing expectations for efficiency and timeliness~\cite{Li2023intro}~\cite{Han2022intro}. Among these, the \textbf{Cooperative Air-Ground Delivery} system has emerged as a promising solution, leveraging the synergy between Unmanned Aerial Vehicles (UAVs) and ground carriers~\cite{Gao2025rel}. As illustrated in \textbf{Figure \ref{fig:intro}(a)}, UAVs offer high-speed, direct aerial transport unaffected by ground traffic but are constrained by battery capacity and payload limits~\cite{Wu2025intro}. Conversely, ground carriers possess superior endurance and payload capacity yet suffer from road network congestion. 
In that sense, the dispatching system are required to solving a sequential decision-making problem~\cite{Zong2023int} with heterogeneous transportation mode and complex spatio-temporal constrains, where the system must continuously assign orders to suitable vehicles to maximize the overall delivery quality.


Order dispatching is often formulated as a dynamic combinatorial optimization problem~\cite{Chen2023int}, and existing solutions typically employ exact solvers or heuristic algorithms~\cite{Nina2021intro}. While exact solvers ensure optimality, they suffer from prohibitive computational costs in large-scale settings~\cite{Yoshua2021intro}. Conversely, heuristics, such as greedy approaches or meta-heuristics, offer computational efficiency but are often myopic~\cite{Fan2024intro}, failing to optimize long-term utility in dynamic environments~\cite{Wen2023int}.
In contrast, RL-based methods have demonstrated superior capability in capturing temporal dependencies and maximizing cumulative rewards~\cite{Yang2024rel}. However, directly applying existing RL frameworks faces two significant limitations:

\par \textbf{1) Heterogeneity in diverse agents}. 
As in Figure \ref{fig:intro}(b), 
most existing RL-based dispatching approaches treat all vehicles with unified state representations and policy networks~\cite{Zong2024int}~\cite{Zong2025int} (a.k.a. homogeneous). 
However, UAVs and carriers are inherently heterogeneous in terms of mobility logic (3D flight vs. 2D road network), energy constraints (battery vs. fuel), and distance metrics (Euclidean vs. Route distance)~\cite{Li2025intro}. 
A unified embedding fails to capture these distinct characteristics, increasing the risk of suboptimal decisions, such as assigning long-distance orders to UAVs with insufficient battery or time-critical orders to carriers in congested zones.

\textbf{2) Scalability bottleneck} as the vehicle size expands. As in Figure \ref{fig:intro}(c), in real-world logistics, the number of vehicles and orders can be substantial. 
Traditional centralized RL architectures face an exponential explosion in the action space as the number of UAVs and carriers increases~\cite{Yang2024rel}. 
Similarly, Multi-Agent RL paradigms, restricted by their flat coordination structures, also struggle to manage the combinatorial explosion of the joint action space in large-scale heterogeneous fleets~\cite{Pateria2021intro}.
This curse of dimensionality leads to inefficient exploration, slow convergence during training, and unacceptable latency during inference, hindering the deployment of such models in large-scale, real-time response scenarios~\cite{Sadeghi2022intro}.

To tackle these challenges, we propose a framework called \underline{H}ierar-
chical \underline{R}einforcement \underline{L}earning for cooperative \underline{A}ir-\underline{G}round delivery (HRL4AG). To address the first limitation, HRL4AG employs mode-specific \emph{Worker Agents} that utilize distinct embedding layers to encode the heterogeneous state information and distance metrics of UAVs and carriers separately. This allows the model to learn specialized scoring functions for feasible order-delivery (O-D) pairs tailored to each delivery mode. To address the second limitation, we introduce a high-level \emph{Manager Agent} that decides whether an order should be assigned to the UAV fleet or the carrier fleet. By decomposing the massive search space into a hierarchical structure, HRL4AG reduces the action space by more than half per decision step. Furthermore, we design a novel internal reward mechanism to improve training stability and exploration efficiency.

Our contributions are summarized as follows:
\begin{itemize}[leftmargin=*]
    \item \textit{Cooperative Air-Ground Delivery Framework}: To the best of our knowledge, we are the first to propose a framework from dispatching to execution for heterogeneous cooperative air-ground delivery, considering both road-network constraints and low-altitude flight characteristics.
    \item \textit{Hierarchical Decision-Making with Internal Incentives}: We develop the HRL4AG algorithm 
    which addresses the limitations of \textit{heterogeneity gaps} and \textit{scalability bottlenecks}. We further propose an internal incentive mechanism combined with reward shaping to enhance the effectiveness and efficiency of the policy.
    \item \textit{Comprehensive empirical evidence and Platform Evaluation}: 
    We conduct extensive experiments on two real-world datasets and deploy our algorithms on an evaluation platform. The results demonstrate that HRL4AG significantly outperforms representative baselines, \textit{improving the delivery success rate by up to 26\% while achieving an 80-fold increase in computational efficiency}.
\end{itemize}

\begin{figure*}[!t] 
    \centering
    \includegraphics[width=0.9\textwidth]{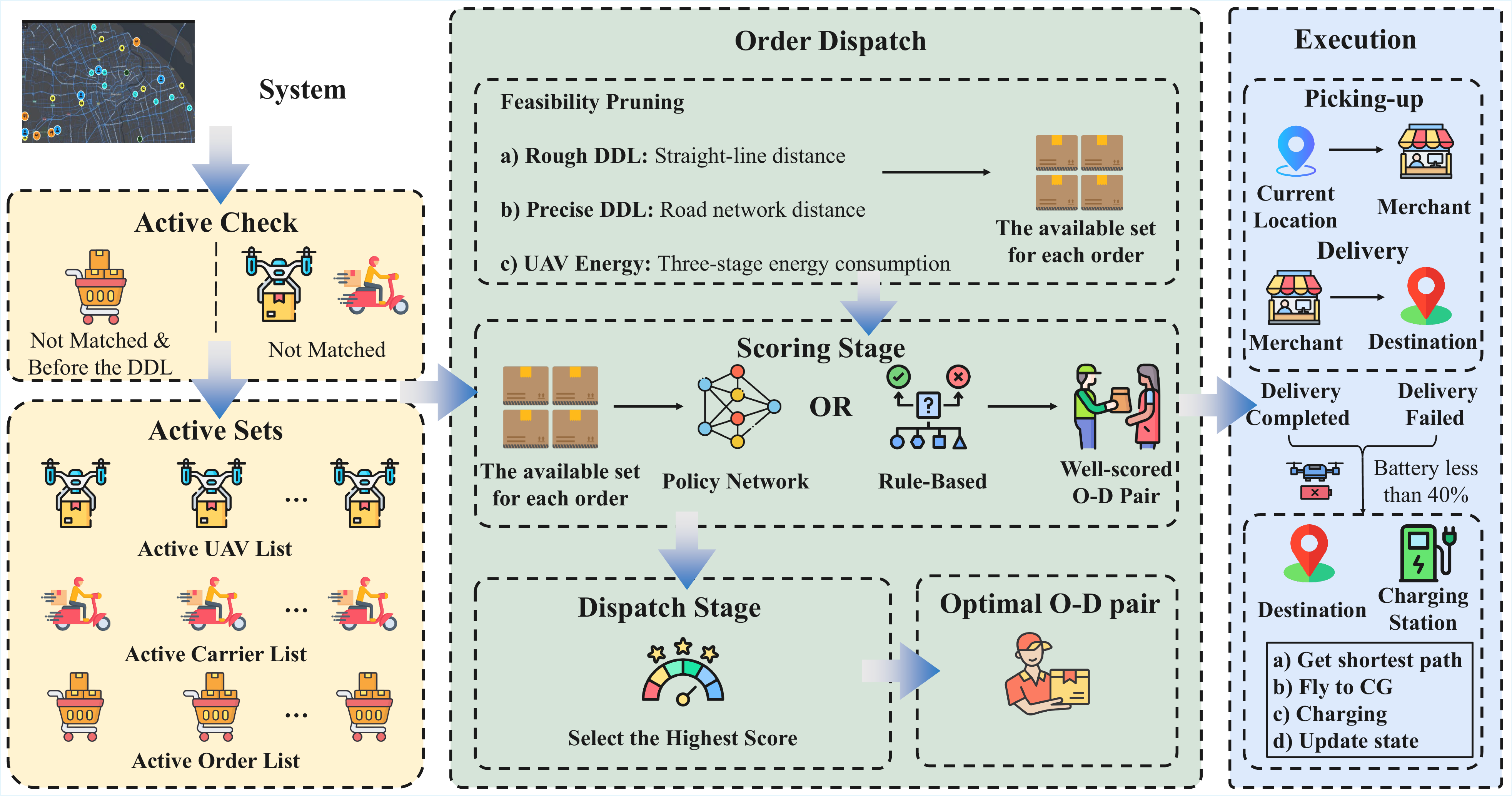}
    \vspace{-0.5em}
    \caption{Cooperative Air-Ground Delivery Framework.}
    \label{fig:workflow}
    \vspace{-1em}
\end{figure*}

\section{Preliminary}
\label{sec:preliminary}

Let $\mathcal{P}$, $\mathcal{U}$, and $\mathcal{V}$ denote the sets of pending orders, UAVs, and carriers, respectively.
We define two binary decision variables: $x_{p,u}$ indicates whether the order $p$ is assigned to the UAV $u$, and $y_{p,v}$ indicates whether the order $p$ is assigned to the carrier $v$.
The problem is formulated as an Integer Linear Programming model:
{\footnotesize
\begin{align}
\mathcal{P}_{opt}: \;& \max \sum_{p \in \mathcal{P}} \left( \sum_{u \in \mathcal{U}} x_{p,u} + \sum_{v \in \mathcal{V}} y_{p,v} \right) \label{eq:objective} \\
s.t. \;\; 
& C_1: x_{p,u}, y_{p,v} \in \{0, 1\}, \quad \forall p, u, v \tag{\ref{eq:objective}{a}} \\
& C_2: \sum_{u \in \mathcal{U}} x_{p,u} + \sum_{v \in \mathcal{V}} y_{p,v} \leq 1, \quad \forall p \in \mathcal{P} \tag{\ref{eq:objective}{b}} \\
& C_3: y_{p,v} \leq S(v, p), \quad \forall p \in \mathcal{P}, v \in \mathcal{V} \tag{\ref{eq:objective}{c}} \\
& C_4: x_{p,u} \leq S(u, p), \quad \forall p \in \mathcal{P}, u \in \mathcal{U} \tag{\ref{eq:objective}{d}}
\end{align}
}

\noindent where $C_1$ ensures the binary nature of decision variables. $C_2$ guarantees that each order is served by at most one vehicle (either a UAV or a carrier).
Constraints $C_3$ and $C_4$ enforce the feasibility of the assignment, where $S(\cdot, \cdot)$ is a binary indicator representing whether a vehicle has the capability to deliver an order.
Specifically, for carriers ($C_3$), $S(v, p)=1$ only if the order deadline (DDL) can be met under ground traffic conditions.
For UAVs ($C_4$), $S(u, p)=1$ requires satisfying both the DDL and the \emph{Safe Return Energy Constraint}:
\begin{equation}
    E_{curr}^{u} \geq \eta \cdot (d(loc_u, loc_p) + d(loc_p, \mathcal{C}_{nearest})),
\end{equation}
where $E_{curr}^{u}$ is the current battery level, $\eta$ is the energy consumption rate per unit distance, $d(\cdot)$ denotes the distance, and $\mathcal{C}_{nearest}$ is the nearest charging station to the order's destination.

\section{Cooperative Air-Ground Delivery Framework}
\label{sec:workflow}


To bridge the gap between theoretical algorithms and real-world deployment, we propose a framework that orchestrates the lifecycle of cooperative air-ground delivery~\cite{Wang2025com}. As shown in \textbf{Figure~\ref{fig:workflow}}, the system operates through three phases at each decision epoch:

\noindent \textbf{Phase 1: Active Check.}
This phase constructs the candidate pools for the current step. The system filters for \textbf{active orders} (unserved requests within valid time windows) and \textbf{idle vehicles} (UAVs and carriers not currently occupied). Only these valid entities are passed to the subsequent dispatching phase~\cite{Guo2025com}.

\noindent \textbf{Phase 2: Order Dispatch.}
This core phase matches orders with suitable vehicles via  \emph{Feasibility Pruning} and \emph{Assignment}.

\textbf{Feasibility Pruning.} 
To reduce the search space and ensure physical viability, we apply distinct constraint filters:
\begin{itemize}[leftmargin=*]
    \item \textbf{Dual-level DDL Check:} We first apply a \emph{Euclidean-based} deadline check to filter obvious mismatches for all vehicles. For carriers specifically, a rigorous \emph{Road Network-based} verification follows to account for traffic tortuosity and real-world constraints.
    \item \textbf{UAV Energy Safety Check:} For UAVs, we enforce a critical \emph{Safe Return} constraint~\cite{Qi2024com}. A match is valid only if the UAV's battery supports the full loop: \emph{Current$\to$Pickup$\to$Delivery$\to$Nearest Charging Station}, ensuring mission safety.
\end{itemize}

\textbf{Scoring and Assignment.} 
Feasible pairs are then evaluated by a scoring function (derived from our RL policy or heuristics) and assigned via a greedy matching strategy to maximize global utility, ensuring one-to-one mapping constraints~\cite{Wu2024com}.

\noindent \textbf{Phase 3: Execution and State Transition.}
Upon assignment, the matched vehicles transition to the \emph{working} state and physically perform the delivery task~\cite{Guo2023com}. Crucially, the system handles delivery state transitions heterogeneously based on vehicle type:
\begin{itemize}[leftmargin=*]
    \item \textbf{Carriers:} Upon completion, carriers immediately become idle at the destination to await new orders~\cite{Hao2024com}.
    \item \textbf{UAVs:} The system triggers a residual battery check. If energy falls below a safety threshold (e.g., 40\%), the UAV is routed to the nearest charging station and remains unavailable until fully replenished; otherwise, it becomes idle at the current location.
\end{itemize}

\section{Methodology}

\subsection{Order Dispatch Modeled as Markov Decision Process}

We formulate the cooperative air-ground delivery problem as an Order-Delivery oriented Markov Decision Process (OD-MDP), modeled by the tuple $\langle \mathcal{T}, \mathcal{S}, \mathcal{A}, P, R, \gamma \rangle$. In this framework, the entire system, including the central dispatching platform and the heterogeneous fleet, is viewed as a unified learning agent. The definitions of each component are detailed as follows.

\noindent\textbf{State Space $\mathcal{S}$.}
We construct the global state $s_t \in \mathcal{S}$ as a fixed-length single vector composed of four concatenated segments. All features are normalized to $[0, 1]$ to improve training stability.

\begin{itemize}[leftmargin=*]
    \item \emph{Global Order Info ($s^{order}_t$):} Features of the active orders. Each order $o_i$ is encoded as $[l^{start}_i, l^{end}_i, \tau^{start}_i, \tau^{ddl}_i]$, which means the locations, start time, and deadline.
    \item \emph{UAV Fleet ($s^{uav}_t$):} States of $N_u$ UAVs. Each UAV $v_j$ is represented by $[l^{curr}_j, \nu_j, \delta_j, e_j]$, denoting location, speed, availability ($\delta_j \in \{0,1\}$), and battery ratio, respectively.
    \item \emph{Carrier Fleet ($s^{car}_t$):} States of $N_c$ carriers, represented by $[l^{curr}_k, \nu_k, \delta_k]$. Unlike UAVs, carriers do not include energy constraints in their state as they are assumed to have sufficient fuel for operations.
    \item \emph{System Statistics ($s^{sys}_t$):} An 8-dimensional vector tracking macroscopic dynamics: timestep, order ratio, matching rate, fleet utilization, UAV failure rate, charging frequency, and UAV proportion.
\end{itemize}

Formally, the global state is: $s_t = s^{order}_t \oplus s^{uav}_t \oplus s^{car}_t \oplus s^{sys}_t$.

\noindent\textbf{Action Space $\mathcal{A}$.}
Given the dynamic number of active orders and available vehicles, the feasible action space varies across time steps. We define a unified action space with a maximum capacity and employ an action masking mechanism to handle invalid pairs. Under our setting, the action space is decomposed into two levels:
\begin{itemize}[leftmargin=*]
    \item \emph{Manager Action $a^m_t$:} For each active order $o_i$, the manager outputs a binary selection $a^m_{t,i} \in \{0, 1\}$, directing the order to either the UAV fleet or the Carrier fleet. This serves as a high-level goal $g_t$.
    \item \emph{Worker Action $a^w_t$:} Based on the manager's directive, the workers assign a continuous score $w_{i,j} \in [0, 1]$ to each feasible O-D pair $(o_i, v_j)$. The final dispatching decision is made by greedily selecting pairs with the highest scores. The dimension of $a^w_t$ is bounded by $N_{max} \times M$, where $N_{max}$ is the maximum concurrent orders. Invalid or pruned pairs are masked with $-\infty$ scores.
\end{itemize}

\begin{figure}[!t]
\centering
\includegraphics[width=0.46\textwidth]{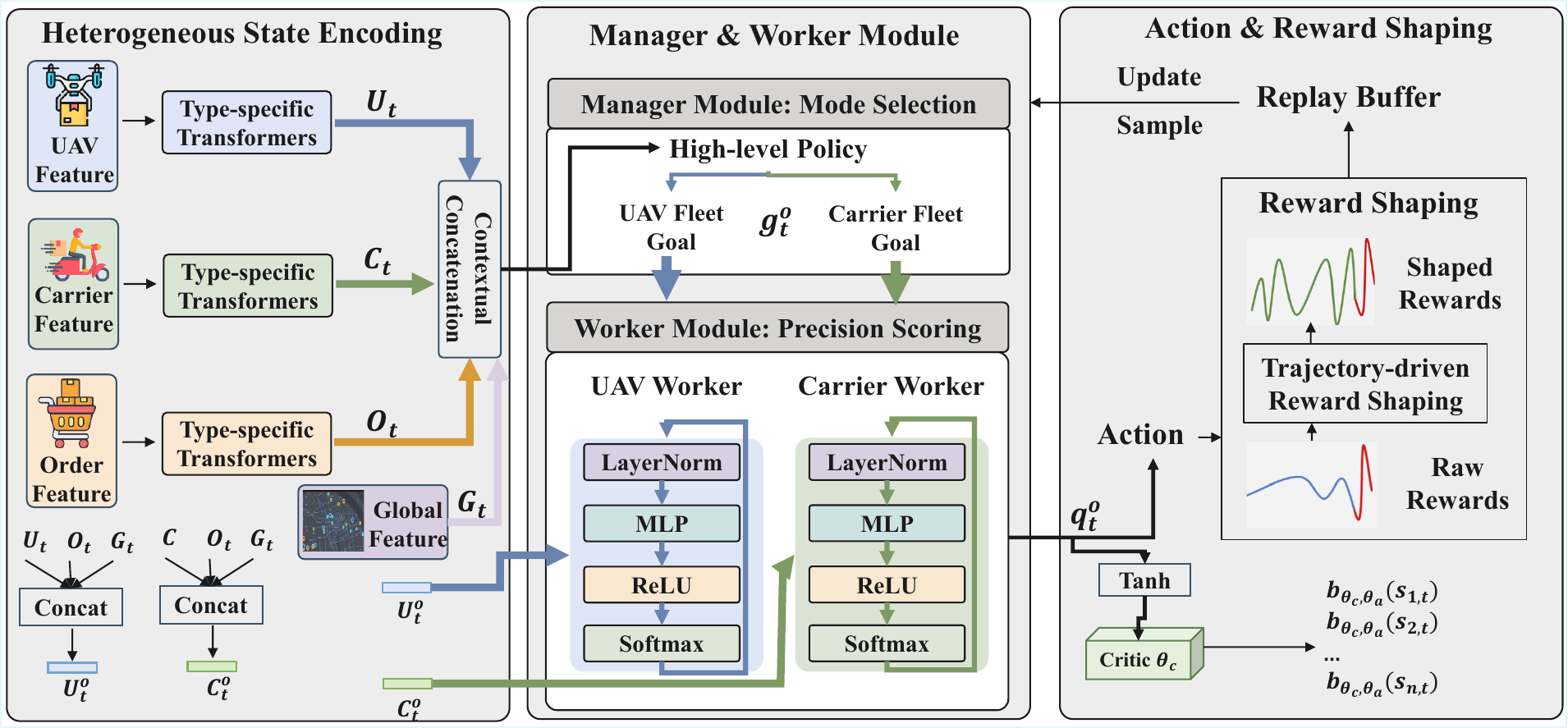}
\vspace{-0.5em}
\caption{Architecture of HRL4AG.}\label{fig:architecture}
\vspace{-1.5em}
\end{figure}

\noindent\textbf{Transition $P$.}
The transition probability $P(s_{t+1}|s_t, a_t)$ describes the system dynamics~\cite{Wang2021com}. Upon executing the dispatching action $a_t$, the environment updates the locations and status of assigned vehicles (e.g., from idle to busy), decreases the battery of flying UAVs, and removes completed or expired orders from the active set.

\noindent\textbf{Reward Function $R$.}
To address the credit assignment problem in cooperative tasks, we design a hierarchical reward structure:
\begin{itemize}[leftmargin=*]
    \item \emph{Extrinsic Reward $r^{ex}_t$:} The manager receives the global extrinsic reward, which aims to maximize long-term service performance. It is defined as a weighted sum of successfully matched, picked-up, and delivered orders at time $t$, augmented by \emph{reward shaping} to alleviate sparsity (detailed in Section \ref{sec:shaping}).
    \item \emph{Intrinsic Reward $r^{in}_t$:} The workers receive a hybrid reward combining the extrinsic signal and an intrinsic reward generated by the Manager. $r^{in}_t$ incentivizes the workers to align their scoring behavior with the Manager's group-selection goals (detailed in Section \ref{sec:manager} and Section\ref{sec:worker}).
\end{itemize}

\subsection{Overview of HRL4AG}


The primary objective of HRL4AG is to learn a joint policy that coordinates heterogeneous fleets to maximize the long-term delivery throughput. As illustrated in \textbf{Figure~\ref{fig:architecture}}, the framework operates through three tightly coupled modules:
\begin{itemize}[leftmargin=*]
    \item \textbf{Heterogeneous State Encoding.} 
    To tackle the inputs from distinct domains, this module employs type-specific transformers to extract permutation-invariant features for UAVs, carriers, and orders. These features are then fused with global context to construct representations for different levels of the hierarchy.
    
    \item \textbf{Manager \& Worker Module.} 
    This core module decomposes the intractable joint action space. A high-level \emph{Manager} first determines the optimal transportation mode (Air vs. Ground) for each order based on global states. Conditioned on this goal, low-level \emph{Workers} then compute fine-grained matching scores for specific O-D pairs within the selected mode, executing the final dispatch via a greedy assignment strategy.
    
    \item \textbf{Trajectory-driven Reward Shaping.} 
    To address the sparse feedback inherent in long-horizon delivery tasks, this module constructs a dense reward signal following the order's lifecycle~\cite{Wang2023com}. It guides the agents by evaluating intermediate milestones (e.g., matching, pickup) and execution efficiency, ensuring stable convergence towards the global objective.
\end{itemize}
During training, agents interact with the environment and get experience. We sample mini-batches from the experience and refine the strategy using DDPG~\cite{Quang2019med} until convergence.

\subsection{Heterogeneous State Encoding}
\label{sec:state}

To handle \emph{the variable number of vehicles} and maintain \emph{permutation invariance} within each homogeneous fleet, we employ three parallel transformer encoders (without positional embeddings) to extract features for UAVs, carriers, and orders, denoted as $\mathbf{U}_t, \mathbf{C}_t, \text{and } \mathbf{O}_t$, respectively~\cite{Dai2023com}. The omission of positional encoding ensures that the learned embeddings rely solely on entity attributes and interactions rather than arbitrary input order~\cite{Wu2025com}~\cite{lyu2026ts}~\cite{lyu2026occamvts}.

To support the hierarchical decision-making, we fuse these entity embeddings with the global system context $\mathbf{G}_t$ to construct distinct inputs for different levels. 
For the low-level workers, we generate mode-specific contexts to evaluate internal matching feasibility:
\begin{equation}
    \mathbf{U}^{o}_{t} = \mathbf{U}_t \oplus \mathbf{O}_t \oplus \mathbf{G}_t, \quad 
    \mathbf{C}^{o}_{t} = \mathbf{C}_t \oplus \mathbf{O}_t \oplus \mathbf{G}_t.
\end{equation}
For the high-level manager, we concatenate all features to form a comprehensive unified representation~\cite{Habib2025com}:
\begin{equation}
    \mathbf{Z}^{o}_{t} = \mathbf{U}_t \oplus \mathbf{C}_t \oplus \mathbf{O}_t \oplus \mathbf{G}_t,
\end{equation}
where $\oplus$ denotes the concatenation operation. These representations ($\mathbf{Z}^{o}_{t}, \mathbf{U}^{o}_{t}, \mathbf{C}^{o}_{t}$) serve as the inputs for the subsequent parts.

\subsection{Manager Agent: High-Level Mode Selection}
\label{sec:manager}

The manager agent operates at the upper level of the hierarchy, acting as a global strategist to decompose the massive joint action space~\cite{Nayyar2025com}. Its core responsibility is to determine the optimal transportation mode (i.e., UAV fleet or Ground Carrier fleet) for each active order based on the macroscopic environment, without getting entangled in specific vehicle-level matching details.

Based on the global order-conditioned representation $\mathbf{Z}^{o}_{t}$ derived in Section~\ref{sec:state}, the Manager employs a policy network $\pi_m$ (composed of LayerNorm and an MLP) to generate a high-level goal:
\begin{equation}
\mathbf{h}^{o}_{t}=\mathrm{MLP}_{m}\!\left(\mathrm{LN}(\mathbf{Z}^{o}_{t})\right),\qquad
\mathbf{g}^{o}_{t}=\mathrm{Softmax}(\mathbf{h}^{o}_{t}),
\end{equation}
where $\mathbf{g}^{o}_{t}\in[0,1]^{2}$ represents the distribution over the two modes. This distribution serves as the \emph{intrinsic goal} passed to the worker, guiding lower-level agents to prioritize the selected candidate set.

To maximize long-term throughput, the manager is trained to optimize the \textbf{global extrinsic reward} $r^{ex}_t$. However, relying solely on sparse delivery completion signals often leads to slow convergence. Therefore, instead of using raw statistics directly, the manager optimizes a dense, trajectory-based shaped reward derived from the system's objective function, which is presented in Section~\ref{sec:shaping}. 

\subsection{Worker Agent: Low-Level Specific Execution}
\label{sec:worker}

The worker agent operates at the lower level, responsible for the precise execution of order dispatching. Conditioned on the manager's goal, the worker evaluates the feasibility of specific vehicles within the selected mode and generates matching scores.

\subsubsection{Policy Design}
For a specific order $o$, the Worker receives two inputs: (i) the mode-specific representations ($\mathbf{U}^{o}_{t}$ for UAV candidates or $\mathbf{C}^{o}_{t}$ for Carrier candidates), and (ii) the goal vector $\mathbf{g}^{o}_{t}$ from the manager. To effectively utilize the manager's guidance, we design a gating mechanism that fuses the goal into the vehicle features:
\begin{equation}
    \mathbf{q}^{o}_{t} = \mathrm{ReLU} \left( \phi(\mathbf{U}^{o}_{t}, \mathbf{C}^{o}_{t}) \oplus \psi(\mathbf{g}^{o}_{t}) \right),
\end{equation}
where $\phi(\cdot)$ is the feature encoder for specific vehicles, $\psi(\cdot)$ projects the manager's goal into the latent feature space, and $\oplus$ denotes concatenation. The final scoring action $a_{t, (o, v)}$ for an O-D pair $(o, v)$ is generated by a bounded output layer:
\begin{equation}
    a_{t, (o, v)} = \tanh(\mathrm{MLP}_{w}(\mathbf{q}^{o}_{t})).
\end{equation}
These scores are used to rank candidate vehicles, and the system greedily selects the highest-scored pairs for execution~\cite{Ling2025com}.

\subsubsection{Hybrid Reward Mechanism}
A core challenge in hierarchical RL is the credit assignment problem, the worker needs distinct feedback on how well it followed the manager's instructions. To address this, we introduce an \textbf{Intrinsic Reward} mechanism.
The worker is trained to maximize a \emph{Hybrid Reward} $r^{w}_t$:
\begin{equation}
    r^{w}_t = r^{ex}_t + \alpha \cdot r^{in}_t,
\end{equation}
where $r^{ex}_t$ is the shared global reward that ensuring the worker cares about the final result, and $r^{in}_t$ is the intrinsic reward. We define $r^{in}_t$ as the consistency between the manager's goal and the worker's actual execution. Specifically, if the manager assigns a high probability to the UAV mode, which means that $g^{o}_{t}[\text{UAV}] > 0.5$, and the worker successfully matches the order to a UAV, the worker receives a positive intrinsic reward; otherwise, it receives a penalty. This mechanism, controlled by the coefficient $\alpha$, significantly stabilizing the training of the bi-level policy.

\subsection{Trajectory-Driven Reward Shaping and Optimization}
\label{sec:shaping}

\textbf{Dense Reward Construction.}
Relying solely on the final delivery count creates a sparse supervision signal that hinders the critic's value estimation. To mitigate this sparsity and guide the hierarchical agents toward optimal policies, we construct a dense extrinsic reward $r^{ex}_t$ that explicitly incorporates trajectory milestones:
\begin{equation}
    r^{ex}_t = \underbrace{(\lambda_m N^m_t + \lambda_p N^p_t + \lambda_d N^d_t)}_{\text{Completion Gradient}} + \underbrace{\lambda_r \frac{N^p_t + N^d_t}{\sum_{i=0}^t{N_{i}^{m}}}}_{\text{Execution Rate}} + \underbrace{\lambda_t (1 - \frac{t}{T_{max}})}_{\text{Time Decay}},
\end{equation}
where $N^m_t, N^p_t, N^d_t$ denote the counts of matched, picked-up, and delivered orders, respectively. 
We assign differentiated weights ($\lambda_m=1.5, \lambda_p=3.0, \lambda_d=10.0$) to create a \emph{completion gradient} that pulls the agent towards order fulfillment. The second term incentivizes the actual execution rate to prevent order hoarding, while the third term ($\lambda_t=0.2$) acts as a temporal regularization to discourage procrastination as the deadline approaches.

\noindent\textbf{Hierarchical Optimization Process.}
To maximize the mentioned cumulative reward, we leverage the Deep Deterministic Policy Gradient (DDPG) algorithm to train the proposed bi-level hierarchy.
Training involves optimizing four neural networks simultaneously: the manager actor-critic pair $(\mu_m, Q_m)$ and the worker actor-critic pair $(\mu_w, Q_w)$.
To ensure stability and break temporal correlations, we utilize a shared \emph{experience replay buffer} $\mathcal{D}$ to store transition tuples $(s_t, a^m_t, a^w_t, r^{ex}_t, r^{in}_t, s_{t+1})$.

During updating, we sample batches from $\mathcal{D}$. Crucially, the manager and workers align with their respective objectives: the manager's target $y^m_t$ is calculated using the global shaped reward $r^{ex}_t$, while the workers' target $y^w_t$ utilizes the hybrid internal reward $r^{in}_t$.
The critics are updated by minimizing the mean squared error loss:
\begin{equation}
    \mathcal{L}(\theta_Q) = \mathbb{E}_{\mathcal{D}} \left[ (y_t - Q(s_t, a_t|\theta_Q))^2 \right],
\end{equation}
where $y_t = r_t + \gamma Q'(s_{t+1}, \mu'(s_{t+1})|\theta_{Q'})$ represents the target value.
Correspondingly, the actors are updated via the deterministic policy gradient to maximize their Q-values:
\begin{equation}
    \nabla_{\theta_\mu} J \approx \mathbb{E}_{s_t \sim \mathcal{D}} \left[ \nabla_a Q(s_t, a|\theta_Q)|_{a=\mu(s_t)} \nabla_{\theta_\mu} \mu(s_t|\theta_\mu) \right].
\end{equation}
Finally, we employ \emph{soft updates} for target network parameters ($\theta' \leftarrow \tau\theta + (1-\tau)\theta'$) to ensure smooth convergence of the policy.

\section{Experiments}

\begin{table}[t] 
    \centering
    \caption{Shanghai Dataset v.s. Chengdu Dataset.}
    \label{tab:dataset}
    \vspace{-1em}
    \renewcommand{\arraystretch}{1.2}
    \resizebox{\columnwidth}{!}{
    \begin{tabular}{l c c}
        \toprule
        \textbf{Parameter} & \textbf{Shanghai} & \textbf{Chengdu} \\
        \midrule
        \multirow{1}{*}{Time Span} & \makecell[c]{30 days } & \makecell[c]{2016/11/01 - 2016/11/6 } \\
        \cmidrule(lr){2-3} 
        \multirow{1}{*}{Spatial Range} & \makecell[c]{$(120.87\degree E,31.50\degree N)$ -  \\ $(121.97 \degree E,30.70 \degree N)$} & \makecell[c]{$(103.99\degree \text{E}, 30.76\degree \text{N})$ - \\ $(104.15\degree \text{E}, 30.63 \degree \text{N})$} \\
        Road Network Nodes & 590,249 & 4,658 \\
        Total Orders & 20,664 & 96,376 \\
        Order Density (/node /day) & 0.001 & 3.448 \\
        Avg. Distance of O-D Pairs & \makecell[c]{Euclidean: 31.37 km \\ Road: 33.79 km} & \makecell[c]{Euclidean: 7.39 km \\ Road: 8.34 km} \\
        \bottomrule
    \end{tabular}
    }
\vspace{-1.5em}
\end{table}

In this section, we conduct extensive experiments to evaluate the effectiveness of HRL4AG in addressing the cooperative air-ground delivery problem. The goal of the experiments is to evaluate the performance of HRL4AG under dynamic urban systems and compare it against baselines, including heuristic algorithms and RL methods.

Specifically, our experiments aim to answer the following research questions (RQs):
\begin{itemize}[leftmargin=*]
    \item \textbf{RQ1:} 
    How does HRL4AG perform compared to other baselines in terms of improving dispatch performance and efficiency?
    \item \textbf{RQ2:} How do the key components of HRL4AG contribute to learning effectiveness and efficiency?
    \item \textbf{RQ3:} How can we assess the contribution of UAVs and carriers to the final delivery performance under different strategies?
    \item \textbf{RQ4:}
    How do the differences between HRL4AG and other baselines in order dispatch decisions lead to varying outcomes?
\end{itemize}

\begin{table*}[!t]
  \centering
  \tabcolsep=1.7mm 
  \vspace{-1em}
    \resizebox{\textwidth}{!}{
    {\footnotesize
    \begin{tabular}{c|c|l|ccc|ccc|ccc}
    \toprule
    \multicolumn{1}{c|}{} & \multicolumn{1}{c|}{\multirow{2}{*}{\textbf{Method}}} & \multicolumn{1}{l|}{\multirow{2}{*}{\textbf{Algorithms}}} 
    & \multicolumn{3}{c|}{\textbf{Nums=30}} & \multicolumn{3}{c|}{\textbf{Nums=40}} & \multicolumn{3}{c}{\textbf{Nums=50}} \\
    \cline{4-12}
    & & & \textbf{PN $\uparrow$} & \textbf{DN $\uparrow$} & \textbf{ET(s) $\downarrow$} & \textbf{PN $\uparrow$} & \textbf{DN $\uparrow$} & \textbf{ET(s) $\downarrow$} & \textbf{PN $\uparrow$} & \textbf{DN $\uparrow$} & \textbf{ET(s) $\downarrow$} \\
    \hline
    \multirow{8}{*}{\rotatebox{90}{Shanghai}} &\multirow{3}{*}{Heuristic} 
    & B\&B & 8{\tiny $\pm$2} & 6{\tiny $\pm$3} & \underline{14.51} & 9{\tiny $\pm$1} & 6{\tiny $\pm$1} & \underline{15.01} & 11{\tiny $\pm$3} & 8{\tiny $\pm$1} & \underline{11.71} \\
    & & Greedy  & 6{\tiny $\pm$2} & 4{\tiny $\pm$1} & 16.54 & 7{\tiny $\pm$1} & 6{\tiny $\pm$1} & 20.25 & 10{\tiny $\pm$3} & 8{\tiny $\pm$1} & 22.18 \\
    & & HGR & 24{\tiny $\pm$3} & 20{\tiny $\pm$2} & 15.07 & 25{\tiny $\pm$2} & 20{\tiny $\pm$2} & 17.49 & 29{\tiny $\pm$3} & 25{\tiny $\pm$2} & 20.23 \\
    \cline{2-12}
    &\multirow{4}{*}{RL} 
    & D2SN  & 27{\tiny $\pm$3} & \underline{22{\tiny $\pm$2}} & 25.9 & 31{\tiny $\pm$3} & 24{\tiny $\pm$3} & 28.5 & 33{\tiny $\pm$4} & 27{\tiny $\pm$3} & 29.6 \\
    & & DECO & 25{\tiny $\pm$2} & 21{\tiny $\pm$2} & 23.3 & 29{\tiny $\pm$3} & 23{\tiny $\pm$3} & 25.6 & 34{\tiny $\pm$3} & 23{\tiny $\pm$3} & 28.1 \\
    & & GRC & 25{\tiny $\pm$4} & 18{\tiny $\pm$3} & 30.5 & 28{\tiny $\pm$5} & 22{\tiny $\pm$4} & 35.6 & 31{\tiny $\pm$5} & 23{\tiny $\pm$4} & 37.2 \\
    & & DDPG-JOTOCC & \underline{28{\tiny $\pm$2}} & 21{\tiny $\pm$2} & 20.7 & \underline{34{\tiny $\pm$3}} & \underline{29{\tiny $\pm$3}} & 24.5 & \underline{37{\tiny $\pm$3}} & \underline{31{\tiny $\pm$3}} & 25.6 \\
    \cline{2-12}\cline{2-12}
    \rowcolor{white} 
    & & \cellcolor{background_gray} 
\textbf{HRL4AG (ours)} & \cellcolor{background_gray} 
\textbf{38}{\tiny $\pm$3}  & \cellcolor{background_gray} 
\textbf{30}{\tiny $\pm$3}  & \cellcolor{background_gray} 
\textbf{0.17}  & \cellcolor{background_gray} 
\textbf{41}{\tiny $\pm$4}  & \cellcolor{background_gray} 
\textbf{36}{\tiny $\pm$3}  & \cellcolor{background_gray} 
\textbf{0.2}  & \cellcolor{background_gray} 
\textbf{49}{\tiny $\pm$5}  & \cellcolor{background_gray} 
\textbf{39}{\tiny $\pm$2}  & \cellcolor{background_gray} 
\textbf{0.35}  \\
    \bottomrule
    \toprule
    \multicolumn{1}{c|}{} & \multicolumn{1}{c|}{\multirow{2}{*}{\textbf{Method}}} & \multicolumn{1}{l|}{\multirow{2}{*}{\textbf{Algorithms}}} 
    & \multicolumn{3}{c|}{\textbf{Nums=30}} & \multicolumn{3}{c|}{\textbf{Nums=40}} & \multicolumn{3}{c}{\textbf{Nums=50}} \\
    \cline{4-12}
    & & & \textbf{PN $\uparrow$} & \textbf{DN $\uparrow$} & \textbf{ET(s) $\downarrow$} & \textbf{PN $\uparrow$} & \textbf{DN $\uparrow$} & \textbf{ET(s) $\downarrow$} & \textbf{PN $\uparrow$} & \textbf{DN $\uparrow$} & \textbf{ET(s) $\downarrow$} \\
    \hline
    \multirow{8}{*}{\rotatebox{90}{Chengdu}} &
    \multirow{3}{*}{Heuristic} 
    & B\&B & 44{\tiny $\pm$2} & 43{\tiny $\pm$2} & 13.04 & 50{\tiny $\pm$2} & 48{\tiny $\pm$2} & 15.92 & 50{\tiny $\pm$3} & 49{\tiny $\pm$2} & 18.01  \\
    & & Greedy  & 38{\tiny $\pm$2} & 37{\tiny $\pm$2} & 7.69 & 42{\tiny $\pm$3} & 40{\tiny $\pm$2} & 9.17 & 44{\tiny $\pm$3} & 41{\tiny $\pm$2} & 8.95 \\
    & & HGR & 61{\tiny $\pm$4} & \underline{59{\tiny $\pm$3}} & 9.55 & \underline{70{\tiny $\pm$5}} & \underline{68{\tiny $\pm$4}} & 10.78 & 73{\tiny $\pm$5} & \underline{71{\tiny $\pm$4}} & 13.36 \\
    \cline{2-12}
    &\multirow{4}{*}{RL} 
    & D2SN  & 59{\tiny $\pm$6} & 55{\tiny $\pm$5} & 5.76 & 68{\tiny $\pm$6} & 64{\tiny $\pm$5} & 6.73 & 71{\tiny $\pm$7} & 69{\tiny $\pm$6} & 8.72 \\
    & & DECO & 53{\tiny $\pm$5} & 50{\tiny $\pm$4} & 5.85 & 67{\tiny $\pm$6} & 64{\tiny $\pm$6} & 6.05 & 70{\tiny $\pm$5} & 68{\tiny $\pm$5} & 9.30 \\
    & & GRC & 49{\tiny $\pm$8} & 47{\tiny $\pm$7} & 7.35 & 61{\tiny $\pm$10} & 54{\tiny $\pm$7} & 9.16 & 65{\tiny $\pm$8} & 60{\tiny $\pm$8} & 11.30 \\
    & & DDPG-JOTOCC  & \underline{64{\tiny $\pm$5}} & 57{\tiny $\pm$5} & \underline{2.81} & 70{\tiny $\pm$5} & 67{\tiny $\pm$5} & \underline{3.40} & \underline{75{\tiny $\pm$6}} & 70{\tiny $\pm$5} & \underline{6.55} \\
    \cline{2-12}
    \rowcolor{white} 
    & & \cellcolor{background_gray} 
\textbf{HRL4AG (ours)} 
& \cellcolor{background_gray} 
\textbf{82}{\tiny $\pm$7} 
& \cellcolor{background_gray} 
\textbf{76}{\tiny $\pm$5} 
& \cellcolor{background_gray} 
\textbf{0.35} 
& \cellcolor{background_gray} 
\textbf{89}{\tiny $\pm$6} 
& \cellcolor{background_gray} 
\textbf{85}{\tiny $\pm$5} 
& \cellcolor{background_gray} 
\textbf{0.56} 
& \cellcolor{background_gray} 
\textbf{95}{\tiny $\pm$8} 
& \cellcolor{background_gray} 
\textbf{88}{\tiny $\pm$7} 
& \cellcolor{background_gray} 
\textbf{0.83} \\
    \bottomrule
    \end{tabular}
    }
}
\caption{5-run experimental results on the Shanghai and Chengdu datasets. Higher PN and DN indicate better  performance, while lower ET signifies higher computational efficiency. Bold and underlined digits are the best and the second best values.}
  \label{tab:results}
  \vspace{-2.5em}
\end{table*}

\subsection{Experimental Settings}

\subsubsection{Datasets}
We preprocessed the aBeacon~\cite{Ding2021dat} and Chengdu~\cite{Yang2024rel} datasets and utilized them 
for cooperative air–ground delivery task. The details of the preprocessed two datasets are shown in \textbf{Table \ref{tab:dataset}}.

\begin{itemize}[leftmargin=*]
    \item  \textbf{Shanghai}: 
    The preprocessed dataset includes 20,664 orders in the area of 6400 $km^2$ of Shanghai. 
    We further crawled the road network within the boundary $(120.87\degree E,31.50\degree N)-(121.97 \degree E,30.70 \degree N)$, and mapped the grid-based order locations to their nearest road network nodes. The dataset contains 30 days orders, where each order consists of the order ID, the pick-up time, the delivery time, the merchant, and the destination. The dataset is split into training (70\%), validation (20\%), and test (10\%) sets by day.
    \item \textbf{Chengdu}: 
    This dataset was collected for 6 days, from November 1st to November 6th, 2016. We constructed the road network within the bounding box $(103.99\degree \text{E}, 30.76\degree \text{N}) - (104.15\degree \text{E}, 30.63 \degree \text{N})$. After data cleaning, the dataset comprises 96,376 orders. Similar to the Shanghai dataset, each order includes the order ID, pick-up time, delivery time, merchant, and destination, which we mapped to the nearest road-network nodes. The dataset is split into training (Day 1-4), validation (Day 5), and test (Day 6) sets.
\end{itemize}

\subsubsection{Baselines}


We adapt several state-of-the-art heuristic and RL algorithms \textit{from top-tier venues within the past three years} to our specific setting for a comprehensive comparison.

\noindent \textbf{Heuristic Algorithms:}
\begin{itemize}[leftmargin=*]
    \item \textbf{B\&B:} Utilizes a branch-and-bound approach with the PuLP ILP solver to determine the optimal solution for the ILP problem.
    \item \textbf{Greedy~\cite{Gao2024rel}:} 
    The dispatch priority is determined by the sum of the distance from the UAV or carrier to the order's origin and the distance from the origin to the destination. Assignments are made by selecting pairs with the smallest total distance. 
    \item \textbf{HGR~\cite{Luo2023bas}:}
    This baseline first models the dispatch process as a graph partitioning problem, using minimum-weight perfect matching to group orders into feasible clusters. Assignments are then determined by constructing a minimum spanning forest that connects these order clusters to available UAVs or carriers.
\end{itemize}

\noindent \textbf{RL Algorithms:}
\begin{itemize}[leftmargin=*]
    \item \textbf{D2SN~\cite{Yue2024rel}:} 
    Leveraging a Deep Double Scalable Network with an encoder-decoder architecture, the approach handles variable-sized inputs to directly generate assignment pairs between orders and vehicles, aiming to maximize long-term cumulative rewards.
    \item \textbf{DECO~\cite{Lu2024bas}:} 
    This baseline enables collaborative order transfers between UAVs or carriers. By integrating a graph-based prediction module with a transfer optimization strategy, it dynamically adjusts assignments to enhance global efficiency, actively leveraging agent interactions to handle real-time uncertainties.
    \item \textbf{GRC~\cite{Yang2024rel}:} 
    This baseline revisits traditional value estimation by introducing a supply-demand correction mechanism into the order dispatch process. By dynamically adjusting matching weights based on the real-time scarcity of UAVs or carriers in specific regions, it balances the spatial distribution of resources to mitigate local oversupply and maximize global long-term efficiency.
    \item \textbf{DDPG-JOTOCC~\cite{Bai2025bas}:} 
    Originally designed for UAV-assisted environments, it models the decision-making process as a continuous control problem. It optimizes scheduling decisions, enabling precise control over high-dimensional variables in dynamic settings.
\end{itemize}

\subsubsection{Evaluation Metrics}
We follow previous work~\cite{Gao2024rel} and use pick-up number (PN) and delivery number (DN) to assess performance. In addition, to evaluate the efficiency of baselines and our HRL4AG, we calculate the average duration of each epoch during the inference phase, termed execution time (ET). ET quantifies the decision-making time required by an algorithm to complete a single epoch, serving as a key metric to evaluate computational efficiency. Each method is run 5 times with different random seed settings, and we report the average value along with the variance.

\subsubsection{Implementation Details}
All baseline methods are implemented in Python, with training conducted using PyTorch 2.1.3 on a single NVIDIA RTX A4000 GPU. In terms of the default setting of our HRL4AG, the batch size is set to 32, the initial learning rate of both actor and critic network is set to 0.005, and the hidden size is set to 256. We configure each epoch with a maximum of 50 time steps and initialize UAV energy at 1 kWh. Vehicle speeds are adapted to specific scenarios: for the uncongested Shanghai dataset (sparse orders), UAV and carrier speeds are set to 120 km/h and 90 km/h, while for the congested Chengdu dataset (dense orders), they are set to 60 km/h and 45 km/h. The vehicle number is explored within the range of 30 to 50, where the ratio of UAVs to carriers is 1:1.

\vspace{-1em}
\subsection{Model Comparison (RQ1)}

In this section, we compare the performance of HRL4AG with existing baselines in terms of dispatch performance and computational efficiency, using two real-world datasets as shown in \textbf{Table \ref{tab:results}}. Our HRL4AG consistently outperforms all baselines under varying vehicle numbers $Nums$, achieving an improvement of approximately 18\% to 28\% in delivery number on the Shanghai dataset, and 20\% to 25\% on the Chengdu dataset compared to the second best baselines. This demonstrates the effectiveness of our method for collaborative air-ground delivery.
As the number of vehicles $Nums$ increases, we observe an upward trend in both PN and DN for all baselines across both datasets. This is attributed to the increased dispatch capacity allows a larger proportion of concurrent orders to be fulfilled.

Specifically, on the sparse Shanghai dataset, HRL4AG effectively leverages air-ground heterogeneity. Unlike baselines that struggle with global coordination over a wide area with an ET between 11s to 20s, HRL4AG employs distinct worker agents for UAVs and carriers. This design not only optimizes the O-D pair scoring for each mode to maximize delivery number but also significantly reduces the action search space, maintaining an ultra-low inference time compared to the sluggish performance of baselines.

Conversely, on the dense Chengdu dataset, HRL4AG overcomes the scalability bottleneck where baselines suffer severe computational delays. By utilizing a manager agent to assignment the orders to specific worker so as to narrow the candidate scope, and Euclidean distance Bound for efficient carrier routing, HRL4AG drastically reduces computational overhead. This allows for real-time responsiveness with the ET less than 0.9s, and superior delivery performance with the ET from 80 to 87, proving that high efficiency and effectiveness can be achieved simultaneously.

\subsection{Ablation Study (RQ2)}

\begin{figure}[!t]  
  \centering
  \includegraphics[width=0.95\linewidth]{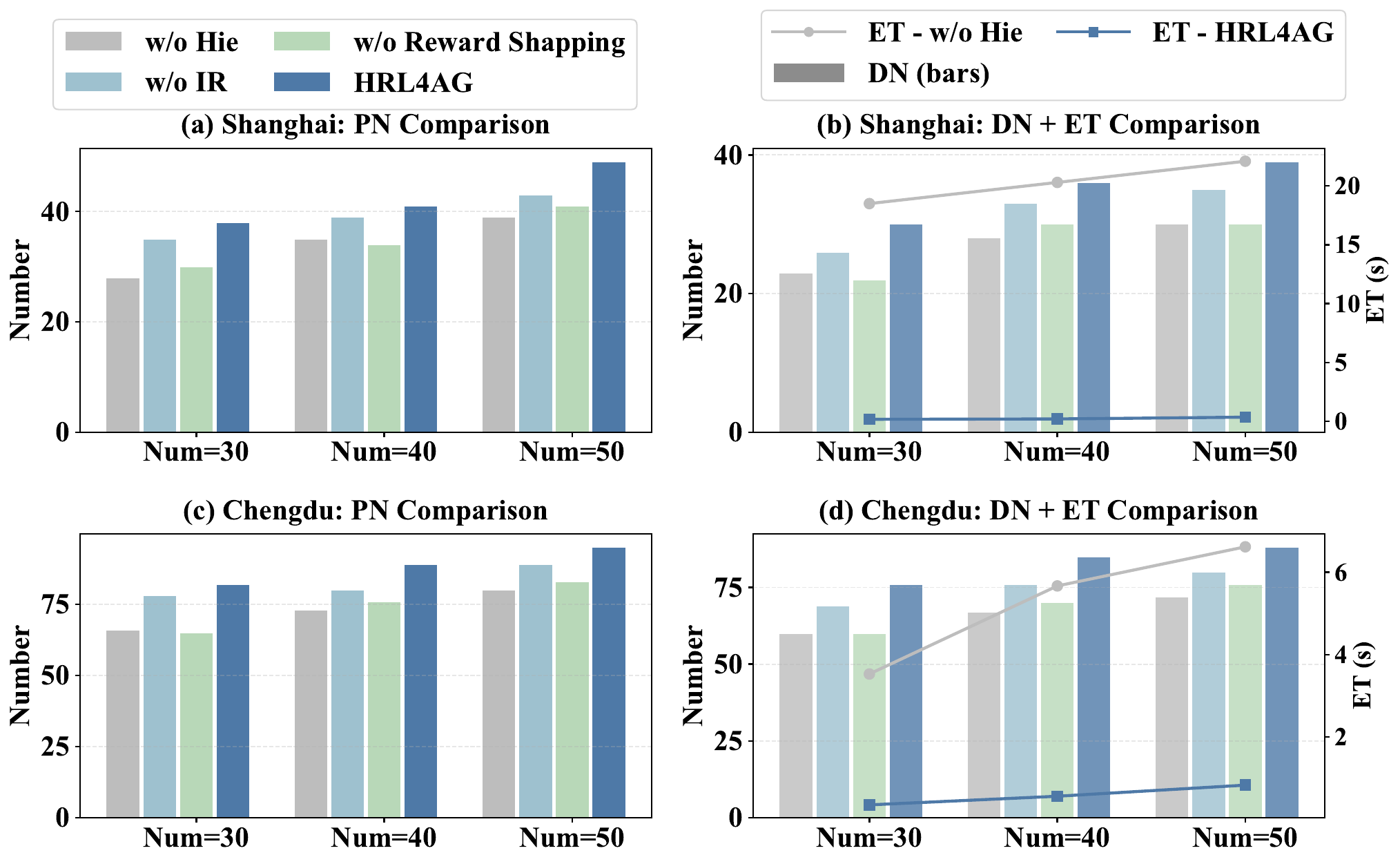}  
  \vspace{-1.5em}
  \caption{Ablation Study Results.}  
  \label{fig:ablation}
  \vspace{-1.5em}
\end{figure}


To comprehensively assess the impact of each component on the overall performance of HRL4AG, we perform ablation studies on both the Shanghai and Chengdu datasets. We systematically compare the full model with three distinct variants: \textbf{w/o Hie} (removing the hierarchical structure), \textbf{w/o IR} (removing the intrinsic reward), and \textbf{w/o RS} (removing the reward shaping). By isolating these modules, we demonstrate how HRL4AG addresses the scalability bottlenecks and heterogeneity gap in urban environments.

\subsubsection{Effects of Hierarchical Architecture.}

To evaluate the necessity of the hierarchical design, we compare HRL4AG with a centralized variant, \textbf{w/o Hie}. In this setting, the high-level manager layer is entirely removed, forcing a single agent to directly control the actions of all UAVs and carriers. As shown in \textbf{Fig.~\ref{fig:ablation}}, removing this structure leads to a decline in computational efficiency, with the Execution Time (ET) drastically increasing from a highly responsive 0.35s to an impractical 22.1s on the Shanghai dataset. Furthermore, the effectiveness metrics (PN and DN) also drop significantly. This severe degradation occurs because the centralized agent struggles with the exponential explosion of the joint action space, suffering heavily from the curse of dimensionality. By contrast, our hierarchical formulation effectively decouples the global dispatching problem into manageable sub-tasks. This confirms the necessity of the hierarchical design in reducing the search space, ensuring both the scalability and the feasibility of real-time decision-making.

\subsubsection{Effects of Reward Mechanisms.}
We investigate our reward design by conducting ablation studies on two key components:
\begin{itemize}[leftmargin=*]
    \item \textbf{Intrinsic Reward (w/o IR)}: 
    This variant removes the intrinsic reward mechanism, forcing both the high-level Manager and the low-level Workers to optimize their policies based solely on the same unified global reward. As shown in \textbf{Fig.~\ref{fig:ablation}}, the absence of IR leads to a noticeable performance drop. This decline highlights the impact of the credit assignment problem in large-scale scheduling. Without intrinsic rewards, individual Workers cannot easily isolate their specific contributions to the global outcome, resulting in noisy gradient updates. 
    \item \textbf{Reward Shaping (w/o RS)}:
    This mechanism is designed to guide agents through the highly sparse-reward environment, where positive feedback is delayed until the final completion of the long-horizon objective. As shown in \textbf{Fig.~\ref{fig:ablation}}, removing reward shaping results in the lowest performance among all reward-related variants.  This clearly demonstrates the contribution of reward shaping in providing continuous and dense gradient signals, thereby preventing agents from collapsing into local optima due to the lack of step-by-step guidance.
\end{itemize}

\subsection{Mode Utilization Analysis (RQ3)}

\begin{figure}[!t]
  \centering
  \begin{subfigure}{0.98\linewidth}
    \centering
    \includegraphics[width=\linewidth]{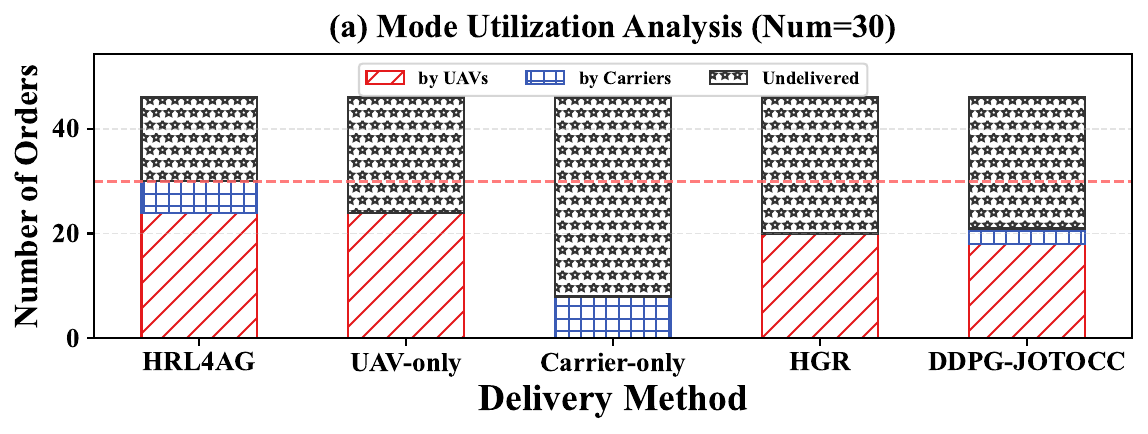}
    \label{fig:mode_30}
  \end{subfigure}
  
  \vspace{-1.5em}
  
  \begin{subfigure}{0.98\linewidth}
    \centering
    \includegraphics[width=\linewidth]{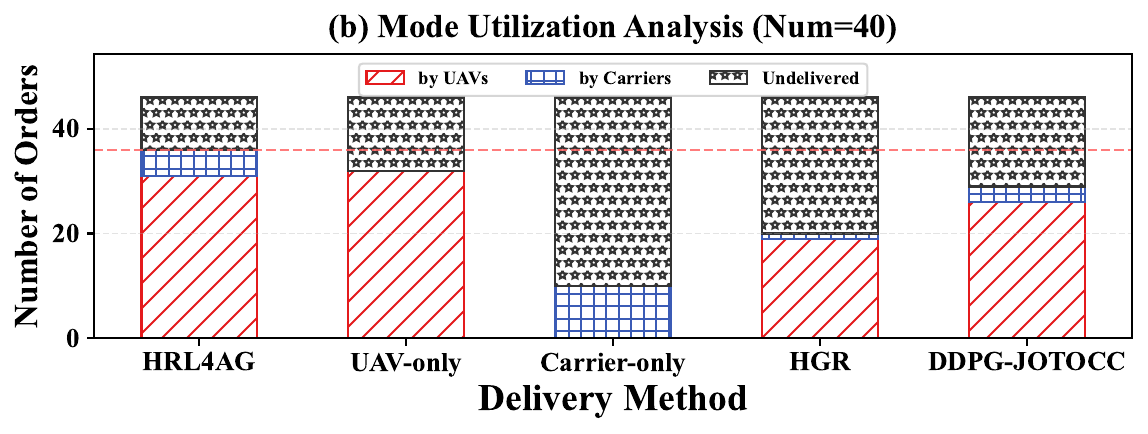}
    \label{fig:mode_40}
  \end{subfigure}
  
  \vspace{-1.5em}
  
  \begin{subfigure}{0.98\linewidth}
    \centering
    \includegraphics[width=\linewidth]{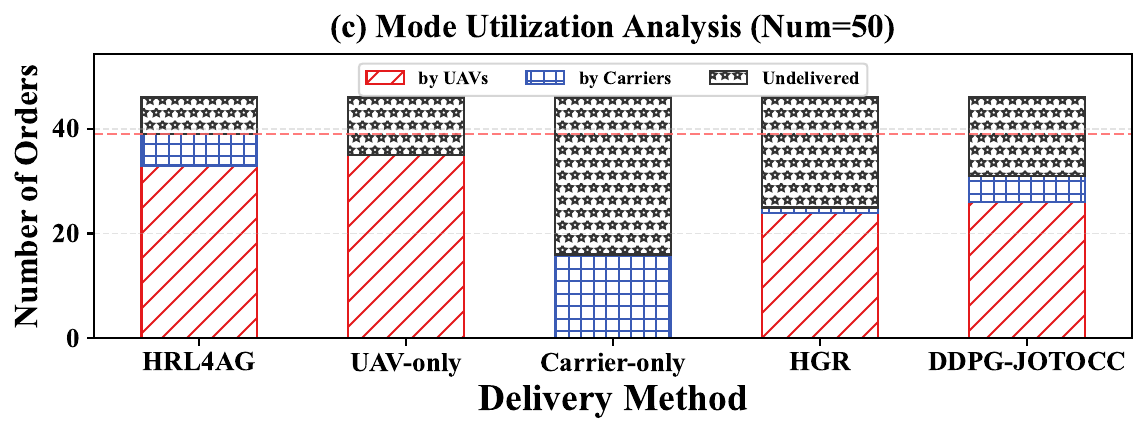}
    \label{fig:mode_50}
  \end{subfigure}
  
  \vspace{-2.5em}
  
  \caption{Mode Utilization Analysis.}
  \label{fig:mode_utilization}
  \vspace{-2em}
\end{figure}


To investigate the specific contributions of different transport modes and validate the inherent necessity of air-ground collaboration, we conduct a comparative analysis under varying fleet sizes ($Nums$). In addition to the top-performing heuristic baseline (\textbf{HGR}) and the multi-agent RL baseline (\textbf{DDPG-JOTOCC}), we introduce two structurally homogeneous variants of our model: \textbf{UAV-only} and \textbf{Carrier-only}. In these extreme variants, the total vehicle size remains strictly constant, but all vehicles are converted to a single type controlled by the HRL4AG logic. This design explicitly isolates the impact of vehicle heterogeneity. To maintain a challenging environment with a high volume of orders relative to $Nums$, we fix the delivery demand ratio at a dense level of 0.5. The comparative results across these different settings are presented in \textbf{Fig.~\ref{fig:mode_utilization}}.


As shown in \textbf{Fig.~\ref{fig:mode_utilization}}, HRL4AG consistently achieves the highest performance across all settings. Notably, while the \textbf{Carrier-only} variant suffers from low speeds and complex traffic constraints, the \textbf{UAV-only} variant, despite the superior speed of drones, also fails to achieve optimal performance. This reveals a critical operational trade-off: relying solely on UAVs exposes the system to severe \textbf{battery constraints}. Although UAVs can bypass road congestion, their limited energy restricts them from completing long-distance or continuous dispatch sequences, leading to higher mission failure rates. 
In contrast, HRL4AG effectively orchestrates a synergy between the two modes. It strategically assigns short-range time-critical requests to UAVs while allocating long-distance or heavy-traffic tasks to Carriers. By leveraging the speed of UAVs and the endurance of carriers, HRL4AG overcomes the limitations of individual modes, achieving the highest delivery rate.

\subsection{Case Study (RQ4)}
To intuitively visualize how HRL4AG differs from baselines in decision-making logic, we developed an \textbf{Cooperative Air-Ground Delivery Simulation Platform}. As shown in Figure 6, this platform allows for the configuration of environmental parameters, including datasets, fleet size, vehicle speeds, and charging station distribution, to visualize the dispatching process in real-time.

We utilized our developed platform to visualize a comparison among three distinct categories of algorithms on the Shanghai dataset. As shown in Figure 6, the visualization panels highlight two critical advantages of HRL4AG. First, in terms of efficiency, the flat baseline RL (DDPG-JOTOCC) exhibits significant decision latency (high Execution Time in Fig. 6b) due to the explosive joint action space, while HRL4AG maintains real-time responsiveness through hierarchical decomposition. Second, regarding decision quality, the heuristic baseline (HGR) exhibits a myopic pattern characterized by low delivery numbers (Fig. 6c), indicating resource depletion from short-sighted assignments. In contrast, HRL4AG outperforms both the heuristic and RL baselines by achieving a balanced conversion rate. This visualization confirms HRL4AG's foresight in avoiding low-quality orders and serves as a white-box debugging tool, enabling developers to trace intermediate dispatch links and iteratively improve reward shaping logic.

\begin{figure}[!t] 
    \centering
    \includegraphics[width=0.45\textwidth]{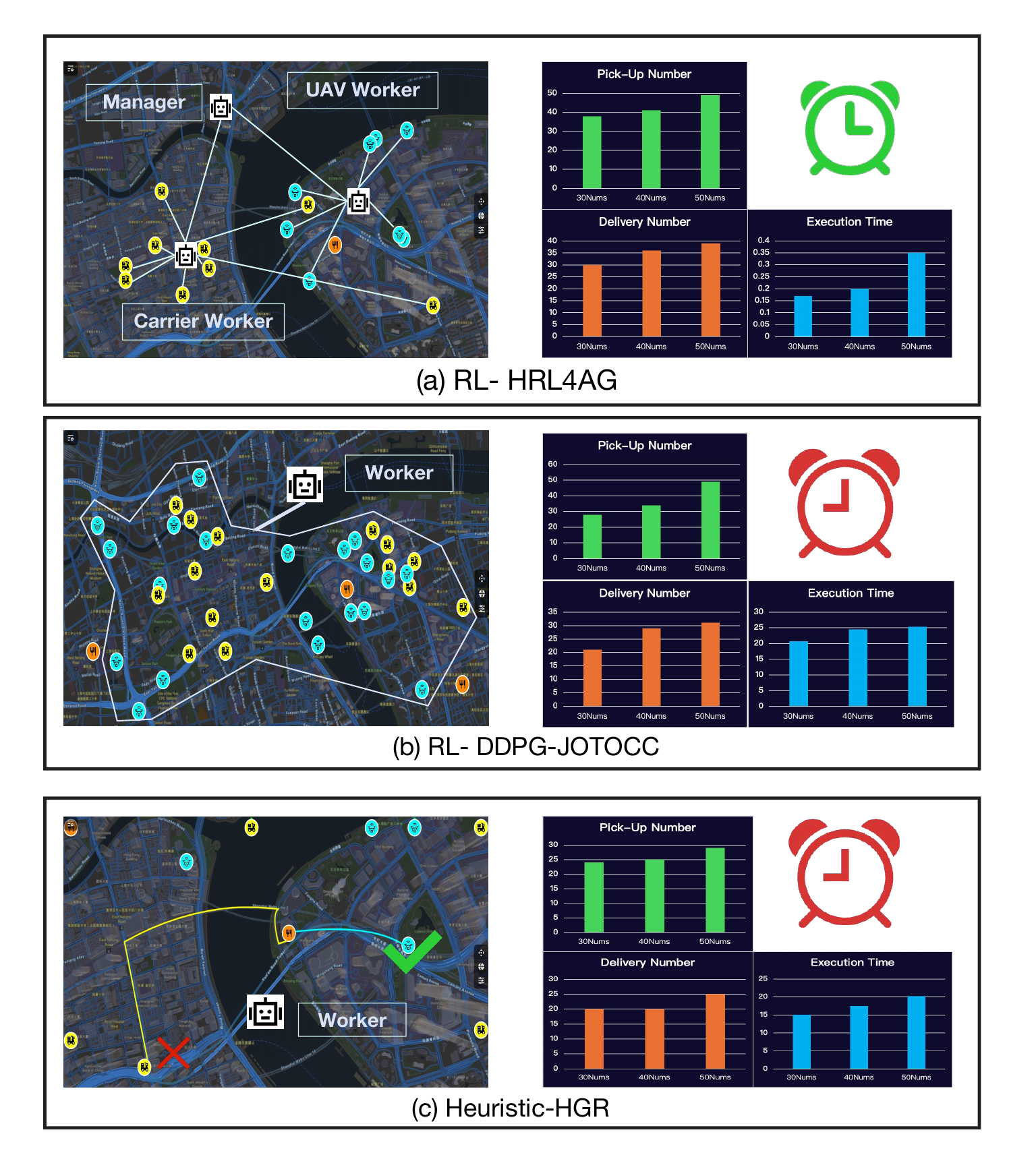}
    \vspace{-1em}
    \caption{Cooperative Air-Ground Delivery Platform.}
    \label{fig:case}
    \vspace{-1.6em}
\end{figure}

\section{Related Works}

\noindent \textbf{Cooperative Air-Ground Delivery.}
Existing research on cooperative air-ground delivery can be broadly categorized into \emph{interactive delivery} and \emph{parallel delivery}. 
Interactive delivery typically treats ground vehicles as mobile launching and retrieval platforms for UAVs to extend their operational range or save energy. For instance, \cite{Das2021rel} models vehicles as carriers and investigates cooperative path planning to synchronize UAV and vehicle trajectories. Other studies explore collaborative delivery for contactless logistics during epidemics~\cite{Wu2022rel} or leverage public transportation (e.g., buses) to assist UAVs~\cite{Pan2021rel, Pan2023rel}. Despite their potential in specific scenarios, these systems require frequent physical synchronization between UAVs and ground vehicles, introducing significant system complexity and operational risks that hinder large-scale deployment.

Consequently, \emph{parallel delivery} has emerged as a more scalable paradigm, where orders are assigned to either UAVs or carriers to be executed independently. Recent works have formulated this as a generalized assignment problem, employing approximation greedy heuristics~\cite{Gao2024rel} or optimization solvers~\cite{Gao2025rel} to determine dispatching schemes. However, most existing methods rely on simplified environmental assumptions, often overlooking the complex road network constraints for ground vehicles and the distinct mobility dynamics between aerial and ground agents. This simplification leads to a gap between theoretical models and real-world efficiency. In contrast, our work explicitly incorporates realistic road-network and flight constraints, proposing a learning-based framework to optimize collaborative efficiency in complex environments.

\noindent \textbf{Reinforcement Learning for Order Dispatch.}
Order dispatching is a core problem in on-demand logistics, often modeled as bipartite matching between orders and vehicles. While traditional combinatorial optimization methods are effective for small-scale static problems, reinforcement learning (RL) has become the dominant approach for large-scale dynamic dispatching due to its ability to optimize long-term returns. 
Related works have proposed various RL architectures. For example, \cite{Yang2024rel} introduced goal-reaching mechanisms to handle credit assignment in large-scale platforms. Other studies utilize hierarchical RL~\cite{Soheil2022rel} or mean-field RL to improve scalability and coordinate agents at both microscopic and macroscopic levels~\cite{Yue2024rel}. Some researches also incorporate fairness considerations into the RL objective function~\cite{Sun2022rel, Jiang2023rel}. 

However, directly transferring these homogeneous RL algorithms to cooperative air-ground delivery is non-trivial. They fail to address the \textit{heterogeneity gap}, as UAVs and carriers possess different states and distance metrics (Euclidean vs. Route). Furthermore, as the number of vehicles grows, classical RL approaches suffer from the \textit{scalability bottleneck} due to the explosion of the joint action space. To handle these limitations, we propose \textbf{HRL4AG}, a hierarchical framework that utilizes mode-specific workers to handle heterogeneity and a high-level manager to decompose the action space, ensuring both effectiveness and efficiency.

\section{Limitations and Ethical Considerations}

While HRL4AG demonstrates strong empirical performance, it has several limitations. 
From a \textbf{modeling} perspective, our objective function primarily prioritizes delivery rate and efficiency, without penalizing UAV energy consumption or carrier detour distances. This may lead to high success rates at the cost of elevated operational overhead. Integrating these factors into a multi-objective reward framework is a promising direction for future refinement. 
Regarding \textbf{system sustainability}, the optimal policy is sensitive to order distributions. Extreme demand patterns, such as highly concentrated short-distance or sparse long-distance demands, may rely on a single transportation mode, leading to uneven fleet wear and potential resource allocation unfairness.


In terms of ethical considerations, our experiments rely on processed public datasets, ensuring no privacy risks. However, real-world deployments must strictly comply with low-altitude airspace regulations. 
Specifically, navigating complex urban environments requires robust safety protocols, including dynamic collision avoidance and strict adherence to temporary no-fly zones.
Furthermore, developers should carefully consider the broader social impact of automation on human couriers. 
Rather than displacing human labor, these autonomous fleets should be designed as collaborative tools that assist workers during peak demands. 

\section{Conclusion \& Future Work}

In this paper, we presented \textbf{HRL4AG}, a hierarchical reinforcement learning framework designed to tackle the cooperative air-ground delivery problem. We identified and addressed two critical limitations that hinder existing approaches: the \textbf{heterogeneity gap} in vehicle dynamics modeling and the \textbf{scalability bottleneck} in large-scale dispatching. To overcome these issues, HRL4AG introduces a Manager-Worker architecture that effectively decomposes the massive action space and utilizes mode-specific embeddings for precise matching, coupled with an internal reward mechanism to ensure training stability. 
Extensive experiments on two real-world
datasets and an evaluation platform demonstrate that HRL4AG
significantly outperforms state-of-the-art baselines, improving the
delivery success rate by up to 26\% while achieving an 80-fold increase in computational efficiency.
In future work, we plan to extend the framework to incorporate stochastic environmental factors, such as weather conditions and uncertain energy consumption.

\section{GenAI Disclosure}
We used a large language model (LLM) to assist with grammar editing of this manuscript. The LLM was not
used to generate scientific content, results, or conclusions. All technical content and claims were produced and verified by the authors.

%
\begin{acks}
This work is supported by the National Natural Science Foundation of China (No.~62402414, No.~62306033, No.~72371217). This work is supported by the Guangdong Basic and Applied Basic Research Foundation (No.~2025A1515011994), Guangdong Provincial Project 2025D03J0014, Guangzhou Municipal Science and Technology Project (No.~2023A03J0011), the Guang-zhou Industrial Information and Intelligent Key Laboratory Project (No.~2024A03J0628), and Guangdong Provincial Key Lab of Integrated Communication, Sensing and Computation for Ubiquitous Internet of Things (No.~2023B1212010007). This work was also supported by the Nansha Key Area Science and Technology Project (No.~2023ZD003, No.~2021JC02X191).
\end{acks}

\clearpage
\bibliographystyle{ACM-Reference-Format}
\bibliography{sample-base}

@String{Computing = "Computing" }

@ARTICLE{Das2021rel,
  author={Das, Dyutimoy Nirupam and Sewani, Rohan and Wang, Junwei and Tiwari, Manoj Kumar},
  journal={IEEE Transactions on Intelligent Transportation Systems}, 
  title={Synchronized Truck and Drone Routing in Package Delivery Logistics}, 
  year={2021},
  volume={22},
  number={9},
  pages={5772-5782},
  doi={10.1109/TITS.2020.2992549}}

@ARTICLE{Wu2022rel,
  author={Wu, Guohua and Mao, Ni and Luo, Qizhang and Xu, Binjie and Shi, Jianmai and Suganthan, Ponnuthurai Nagaratnam},
  journal={IEEE Transactions on Intelligent Transportation Systems}, 
  title={Collaborative Truck-Drone Routing for Contactless Parcel Delivery During the Epidemic}, 
  year={2022},
  volume={23},
  number={12},
  pages={25077-25091},
  doi={10.1109/TITS.2022.3181282}}

@article{Pan2021rel,
author = {Pan, Yan and Li, Shining and Chen, Qianwu and Zhang, Nan and Cheng, Tao and Li, Zhigang and Guo, Bin and Han, Qingye and Zhu, Ting},
title = {Efficient Schedule of Energy-Constrained UAV Using Crowdsourced Buses in Last-Mile Parcel Delivery},
year = {2021},
issue_date = {March 2021},
publisher = {Association for Computing Machinery},
address = {New York, NY, USA},
volume = {5},
number = {1},
url = {https://doi.org/10.1145/3448079},
doi = {10.1145/3448079},
journal = {Proc. ACM Interact. Mob. Wearable Ubiquitous Technol.},
month = mar,
articleno = {28},
numpages = {23}
}

@ARTICLE{Pan2023rel,
  author={Pan, Yan and Chen, Qianwu and Zhang, Nan and Li, Zhigang and Zhu, Ting and Han, Qingye},
  journal={IEEE Transactions on Mobile Computing}, 
  title={Extending Delivery Range and Decelerating Battery Aging of Logistics UAVs Using Public Buses}, 
  year={2023},
  volume={22},
  number={9},
  pages={5280-5295},
  doi={10.1109/TMC.2022.3167040}}

@INPROCEEDINGS{Gao2024rel,
  author={Gao, Junhui and Wang, Qianru and Zhang, Xin and Shi, Juan and Zhao, Xiang and Han, Qingye and Pan, Yan},
  booktitle={2024 IEEE 40th International Conference on Data Engineering (ICDE)}, 
  title={Cooperative Air-Ground Instant Delivery by UAVs and Crowdsourced Taxis}, 
  year={2024},
  volume={},
  number={},
  pages={4153-4166},
  doi={10.1109/ICDE60146.2024.00120}}

@ARTICLE{Gao2025rel,
  author={Gao, Junhui and Wang, Qianru and Zhang, Xin and Shi, Juan and Zhao, Xiang and Liang, Yunji and Guo, Bin and Han, Qingye and Pan, Yan},
  journal={IEEE Transactions on Mobile Computing}, 
  title={Cooperative Air-Ground Instant Delivery by UAVs and Crowdsourced Taxis: Joint UAV Station Deployment and Delivery Scheduling}, 
  year={2025},
  volume={},
  number={},
  pages={1-17},
  doi={10.1109/TMC.2025.3634430}}

@inproceedings{Yang2024rel,
author = {Yang, Zhaoxing and Jin, Haiming and Fan, Guiyun and Lu, Min and Liu, Yiran and Yue, Xinlang and Pan, Hao and Xu, Zhe and Wu, Guobin and Li, Qun and Wang, Xiaotong and Guo, Jiecheng},
title = {Rethinking Order Dispatching in Online Ride-Hailing Platforms},
year = {2024},
isbn = {9798400704901},
publisher = {Association for Computing Machinery},
address = {New York, NY, USA},
url = {https://doi.org/10.1145/3637528.3672028},
doi = {10.1145/3637528.3672028},
booktitle = {Proceedings of the 30th ACM SIGKDD Conference on Knowledge Discovery and Data Mining},
pages = {3863–3873},
numpages = {11},
location = {Barcelona, Spain},
series = {KDD '24}
}

@inproceedings{Yue2024rel,
author = {Yue, Xinlang and Liu, Yiran and Shi, Fangzhou and Luo, Sihong and Zhong, Chen and Lu, Min and Xu, Zhe},
title = {An End-to-End Reinforcement Learning Based Approach for Micro-View Order-Dispatching in Ride-Hailing},
year = {2024},
isbn = {9798400704369},
publisher = {Association for Computing Machinery},
address = {New York, NY, USA},
url = {https://doi.org/10.1145/3627673.3680013},
doi = {10.1145/3627673.3680013},
booktitle = {Proceedings of the 33rd ACM International Conference on Information and Knowledge Management},
pages = {5054–5061},
numpages = {8},
location = {Boise, ID, USA},
series = {CIKM '24}
}

@misc{Soheil2022rel,
      title={Reinforcement Learning in the Wild: Scalable RL Dispatching Algorithm Deployed in Ridehailing Marketplace}, 
      author={Soheil Sadeghi Eshkevari and Xiaocheng Tang and Zhiwei Qin and Jinhan Mei and Cheng Zhang and Qianying Meng and Jia Xu},
      year={2022},
      eprint={2202.05118},
      archivePrefix={arXiv},
      primaryClass={cs.LG},
      url={https://arxiv.org/abs/2202.05118}, 
}

@inproceedings{Sun2022rel,
author = {Sun, Jiahui and Jin, Haiming and Yang, Zhaoxing and Su, Lu and Wang, Xinbing},
title = {Optimizing Long-Term Efficiency and Fairness in Ride-Hailing via Joint Order Dispatching and Driver Repositioning},
year = {2022},
isbn = {9781450393850},
publisher = {Association for Computing Machinery},
address = {New York, NY, USA},
url = {https://doi.org/10.1145/3534678.3539060},
doi = {10.1145/3534678.3539060},
booktitle = {Proceedings of the 28th ACM SIGKDD Conference on Knowledge Discovery and Data Mining},
pages = {3950–3960},
numpages = {11},
location = {Washington DC, USA},
series = {KDD '22}
}

@inproceedings{Jiang2023rel,
author = {Jiang, Lin and Wang, Shuai and Guo, Baoshen and Wang, Hai and Zhang, Desheng and Wang, Guang},
title = {FairCod: A Fairness-aware Concurrent Dispatch System for Large-scale Instant Delivery Services},
year = {2023},
isbn = {9798400701030},
publisher = {Association for Computing Machinery},
address = {New York, NY, USA},
url = {https://doi.org/10.1145/3580305.3599824},
doi = {10.1145/3580305.3599824},
booktitle = {Proceedings of the 29th ACM SIGKDD Conference on Knowledge Discovery and Data Mining},
pages = {4229–4238},
numpages = {10},
location = {Long Beach, CA, USA},
series = {KDD '23}
}

@ARTICLE{Quang2019med,
  author={Quang, Pham Tran Anh and Hadjadj-Aoul, Yassine and Outtagarts, Abdelkader},
  journal={IEEE Transactions on Network and Service Management}, 
  title={A Deep Reinforcement Learning Approach for VNF Forwarding Graph Embedding}, 
  year={2019},
  volume={16},
  number={4},
  pages={1318-1331},
  doi={10.1109/TNSM.2019.2947905}}

@article{Ding2021dat,
author = {Ding, Yi and Liu, Ling and Yang, Yu and Liu, Yunhuai and Zhang, Desheng and He, Tian},
title = {From Conception to Retirement: A Lifetime Story of a 3-Year-Old Wireless Beacon System in the Wild},
year = {2021},
issue_date = {Feb. 2022},
publisher = {IEEE Press},
volume = {30},
number = {1},
issn = {1063-6692},
url = {https://doi.org/10.1109/TNET.2021.3107043},
doi = {10.1109/TNET.2021.3107043},
journal = {IEEE/ACM Trans. Netw.},
month = aug,
pages = {47–61},
numpages = {15}
}

@article{Luo2023bas,
author = {Luo, Kelin and Florio, Alexandre M. and Das, Syamantak and Guo, Xiangyu},
title = {A Hierarchical Grouping Algorithm for the Multi-Vehicle Dial-a-Ride Problem},
year = {2023},
issue_date = {January 2023},
publisher = {VLDB Endowment},
volume = {16},
number = {5},
issn = {2150-8097},
url = {https://doi.org/10.14778/3579075.3579091},
doi = {10.14778/3579075.3579091},
journal = {Proc. VLDB Endow.},
month = jan,
pages = {1195–1207},
numpages = {13}
}

@inproceedings{Lu2024bas,
author = {Lu, Yao and Wang, Shuai and Yang, Yu and Wang, Hai and Guo, Baoshen and Zhang, Desheng and Wang, Shuai and He, Tian},
title = {DECO: Cooperative Order Dispatching for On-Demand Delivery with Real-Time Encounter Detection},
year = {2024},
isbn = {9798400704369},
publisher = {Association for Computing Machinery},
address = {New York, NY, USA},
url = {https://doi.org/10.1145/3627673.3680084},
doi = {10.1145/3627673.3680084},
booktitle = {Proceedings of the 33rd ACM International Conference on Information and Knowledge Management},
pages = {4734–4742},
numpages = {9},
location = {Boise, ID, USA},
series = {CIKM '24}
}

@ARTICLE{Bai2025bas,
  author={Bai, Jingpan and Luo, Jiahui and Chen, Yuan and Tang, Yuming and Jin, Li and Shi, Yan and Yang, Bozhong and Ji, Houling},
  journal={IEEE Internet of Things Journal}, 
  title={The DDPG-Based Joint Optimization of Task Offloading and Content Caching in UAV-Assisted IoV}, 
  year={2025},
  volume={12},
  number={19},
  pages={40330-40346},
  doi={10.1109/JIOT.2025.3588858}}

@article{Li2023intro,
author = {Li, Boyang and Cheng, Yurong and Yuan, Ye and Yang, Yi and Jin, QianQian and Wang, Guoren},
title = {ACTA: Autonomy and Coordination Task Assignment in Spatial Crowdsourcing Platforms},
year = {2023},
issue_date = {January 2023},
publisher = {VLDB Endowment},
volume = {16},
number = {5},
issn = {2150-8097},
url = {https://doi.org/10.14778/3579075.3579082},
doi = {10.14778/3579075.3579082},
journal = {Proc. VLDB Endow.},
month = jan,
pages = {1073–1085},
numpages = {13}
}

@inproceedings{Han2022intro,
author = {Han, Benjamin and Lee, Hyungjun and Martin, S\'{e}bastien},
title = {Real-Time Rideshare Driver Supply Values Using Online Reinforcement Learning},
year = {2022},
isbn = {9781450393850},
publisher = {Association for Computing Machinery},
address = {New York, NY, USA},
url = {https://doi.org/10.1145/3534678.3539141},
doi = {10.1145/3534678.3539141},
booktitle = {Proceedings of the 28th ACM SIGKDD Conference on Knowledge Discovery and Data Mining},
pages = {2968–2976},
numpages = {9},
location = {Washington DC, USA},
series = {KDD '22}
}

@ARTICLE{Wu2025intro,
  author={Wu, Huaming and Tian, Lei and Tang, Huijun and Li, Ruidong and Jiao, Pengfei},
  journal={IEEE Transactions on Intelligent Transportation Systems}, 
  title={Graph Convolutional Reinforcement Learning-Guided Joint Trajectory Optimization and Task Offloading for Aerial Edge Computing}, 
  year={2025},
  volume={26},
  number={10},
  pages={17487-17498},
  doi={10.1109/TITS.2024.3490533}}

@inproceedings{Zong2023int,
author = {Zong, Zefang and Yan, Huan and Sui, Hongjie and Li, Haoxiang and Jiang, Peiqi and Li, Yong},
title = {An AI-based Simulation and Optimization Framework for Logistic Systems},
year = {2023},
isbn = {9798400701245},
publisher = {Association for Computing Machinery},
address = {New York, NY, USA},
url = {https://doi.org/10.1145/3583780.3614732},
doi = {10.1145/3583780.3614732},
booktitle = {Proceedings of the 32nd ACM International Conference on Information and Knowledge Management},
pages = {5138–5142},
numpages = {5},
location = {Birmingham, United Kingdom},
series = {CIKM '23}
}

@article{Chen2023int,
author = {Chen, Jinwei and Zong, Zefang and Zhuang, Yunlin and Yan, Huan and Jin, Depeng and Li, Yong},
title = {Reinforcement Learning for Practical Express Systems with Mixed Deliveries and Pickups},
year = {2023},
issue_date = {April 2023},
publisher = {Association for Computing Machinery},
address = {New York, NY, USA},
volume = {17},
number = {3},
issn = {1556-4681},
url = {https://doi.org/10.1145/3546952},
doi = {10.1145/3546952},
journal = {ACM Trans. Knowl. Discov. Data},
month = feb,
articleno = {33},
numpages = {19}
}

@ARTICLE{Wen2023int,
  author={Wen, Haomin and Lin, Youfang and Hu, Yuxuan and Wu, Fan and Xia, Mingxuan and Zhang, Xinyi and Wu, Lixia and Hu, Haoyuan and Wan, Huaiyu},
  journal={IEEE Transactions on Intelligent Transportation Systems}, 
  title={Modeling Spatial–Temporal Constraints and Spatial-Transfer Patterns for Couriers’ Package Pick-up Route Prediction}, 
  year={2023},
  volume={24},
  number={12},
  pages={13787-13800},
  doi={10.1109/TITS.2023.3301661}}

@article{Zong2024int,
author = {Zong, Zefang and Tong, Xia and Zheng, Meng and Li, Yong},
title = {Reinforcement Learning for Solving Multiple Vehicle Routing Problem with Time Window},
year = {2024},
issue_date = {April 2024},
publisher = {Association for Computing Machinery},
address = {New York, NY, USA},
volume = {15},
number = {2},
issn = {2157-6904},
url = {https://doi.org/10.1145/3625232},
doi = {10.1145/3625232},
journal = {ACM Trans. Intell. Syst. Technol.},
month = mar,
articleno = {32},
numpages = {19}
}

@article{Zong2025int,
author = {Zong, Zefang and Feng, Tao and Wang, Jingwei and Xia, Tong and Li, Yong},
title = {Deep Reinforcement Learning for Demand-Driven Services in Logistics and Transportation Systems: A Survey},
year = {2025},
issue_date = {May 2025},
publisher = {Association for Computing Machinery},
address = {New York, NY, USA},
volume = {19},
number = {4},
issn = {1556-4681},
url = {https://doi.org/10.1145/3708325},
doi = {10.1145/3708325},
journal = {ACM Trans. Knowl. Discov. Data},
month = may,
articleno = {89},
numpages = {42}
}

@article{Li2025intro,
author = {Li, Qingyang and Li, Zexuan and Wang, Qianru and Han, Lei and Cui, Jiangtao and Yu, Zhiwen},
title = {Hierarchical Human-UAV Cooperative Task Allocation for Spatiotemporal Crowdsensing in Disaster Response},
year = {2025},
issue_date = {December 2025},
publisher = {Association for Computing Machinery},
address = {New York, NY, USA},
volume = {9},
number = {4},
url = {https://doi.org/10.1145/3770651},
doi = {10.1145/3770651},
journal = {Proc. ACM Interact. Mob. Wearable Ubiquitous Technol.},
month = dec,
articleno = {188},
numpages = {33}
}

@inproceedings{Sadeghi2022intro,
author = {Sadeghi Eshkevari, Soheil and Tang, Xiaocheng and Qin, Zhiwei and Mei, Jinhan and Zhang, Cheng and Meng, Qianying and Xu, Jia},
title = {Reinforcement Learning in the Wild: Scalable RL Dispatching Algorithm Deployed in Ridehailing Marketplace},
year = {2022},
isbn = {9781450393850},
publisher = {Association for Computing Machinery},
address = {New York, NY, USA},
url = {https://doi.org/10.1145/3534678.3539095},
doi = {10.1145/3534678.3539095},
booktitle = {Proceedings of the 28th ACM SIGKDD Conference on Knowledge Discovery and Data Mining},
pages = {3838–3848},
numpages = {11},
location = {Washington DC, USA},
series = {KDD '22}
}

@article{Nina2021intro,
title = {Reinforcement learning for combinatorial optimization: A survey},
journal = {Computers \& Operations Research},
volume = {134},
pages = {105400},
year = {2021},
issn = {0305-0548},
doi = {https://doi.org/10.1016/j.cor.2021.105400},
url = {https://www.sciencedirect.com/science/article/pii/S0305054821001660},
author = {Nina Mazyavkina and Sergey Sviridov and Sergei Ivanov and Evgeny Burnaev}
}

@article{Yoshua2021intro,
title = {Machine learning for combinatorial optimization: A methodological tour d’horizon},
journal = {European Journal of Operational Research},
volume = {290},
number = {2},
pages = {405-421},
year = {2021},
issn = {0377-2217},
doi = {https://doi.org/10.1016/j.ejor.2020.07.063},
url = {https://www.sciencedirect.com/science/article/pii/S0377221720306895},
author = {Yoshua Bengio and Andrea Lodi and Antoine Prouvost}
}

@misc{Fan2024intro,
      title={Artificial Intelligence for Operations Research: Revolutionizing the Operations Research Process}, 
      author={Zhenan Fan and Bissan Ghaddar and Xinglu Wang and Linzi Xing and Yong Zhang and Zirui Zhou},
      year={2024},
      eprint={2401.03244},
      archivePrefix={arXiv},
      primaryClass={math.OC},
      url={https://arxiv.org/abs/2401.03244}, 
}

@article{Pateria2021intro,
author = {Pateria, Shubham and Subagdja, Budhitama and Tan, Ah-hwee and Quek, Chai},
title = {Hierarchical Reinforcement Learning: A Comprehensive Survey},
year = {2021},
issue_date = {June 2022},
publisher = {Association for Computing Machinery},
address = {New York, NY, USA},
volume = {54},
number = {5},
issn = {0360-0300},
url = {https://doi.org/10.1145/3453160},
doi = {10.1145/3453160},
journal = {ACM Comput. Surv.},
month = jun,
articleno = {109},
numpages = {35}
}

@ARTICLE{Wang2025com,
  author={Wang, Zhenning and Cao, Yue and Jiang, Kai and Zhou, Huan and Kang, Jiawen and Zhuang, Yuan and Tian, Daxin and Leung, Victor C. M.},
  journal={IEEE Communications Surveys \& Tutorials}, 
  title={When Crowdsensing Meets Smart Cities: A Comprehensive Survey and New Perspectives}, 
  year={2025},
  volume={27},
  number={2},
  pages={1101-1151},
  doi={10.1109/COMST.2024.3400121}}

@INPROCEEDINGS{Wang2023com,
  author={Wang, Yu and Wu, Jingfei and Hua, Xingyuan and Liu, Chi Harold and Li, Guozheng and Zhao, Jianxin and Yuan, Ye and Wang, Guoren},
  booktitle={2023 IEEE 39th International Conference on Data Engineering (ICDE)}, 
  title={Air-Ground Spatial Crowdsourcing with UAV Carriers by Geometric Graph Convolutional Multi-Agent Deep Reinforcement Learning}, 
  year={2023},
  volume={},
  number={},
  pages={1790-1802},
  doi={10.1109/ICDE55515.2023.00140}}

@ARTICLE{Dai2023com,
  author={Dai, Zipeng and Liu, Chi Harold and Han, Rui and Wang, Guoren and Leung, Kin K. and Tang, Jian},
  journal={IEEE Transactions on Mobile Computing}, 
  title={Delay-Sensitive Energy-Efficient UAV Crowdsensing by Deep Reinforcement Learning}, 
  year={2023},
  volume={22},
  number={4},
  pages={2038-2052},
  doi={10.1109/TMC.2021.3113052}}

@misc{Wu2025com,
      title={Beyond Regularity: Modeling Chaotic Mobility Patterns for Next Location Prediction}, 
      author={Yuqian Wu and Yuhong Peng and Jiapeng Yu and Xiangyu Liu and Zeting Yan and Kang Lin and Weifeng Su and Bingqing Qu and Raymond Lee and Dingqi Yang},
      year={2025},
      eprint={2509.11713},
      archivePrefix={arXiv},
      primaryClass={cs.LG},
      url={https://arxiv.org/abs/2509.11713}, 
}

@misc{Wu2024com,
      title={MAS4POI: a Multi-Agents Collaboration System for Next POI Recommendation}, 
      author={Yuqian Wu and Yuhong Peng and Jiapeng Yu and Raymond S. T. Lee},
      year={2024},
      eprint={2409.13700},
      archivePrefix={arXiv},
      primaryClass={cs.IR},
      url={https://arxiv.org/abs/2409.13700}, 
}

@misc{Guo2025com,
      title={AgentSense: LLMs Empower Generalizable and Explainable Web-Based Participatory Urban Sensing}, 
      author={Xusen Guo and Mingxing Peng and Xixuan Hao and Xingchen Zou and Qiongyan Wang and Sijie Ruan and Yuxuan Liang},
      year={2025},
      eprint={2510.19661},
      archivePrefix={arXiv},
      primaryClass={cs.AI},
      url={https://arxiv.org/abs/2510.19661}, 
}

@inproceedings{Wang2021com,
author = {Wang, Hao and Liu, Chi Harold and Dai, Zipeng and Tang, Jian and Wang, Guoren},
title = {Energy-Efficient 3D Vehicular Crowdsourcing for Disaster Response by Distributed Deep Reinforcement Learning},
year = {2021},
isbn = {9781450383325},
publisher = {Association for Computing Machinery},
address = {New York, NY, USA},
url = {https://doi.org/10.1145/3447548.3467070},
doi = {10.1145/3447548.3467070},
booktitle = {Proceedings of the 27th ACM SIGKDD Conference on Knowledge Discovery \& Data Mining},
pages = {3679–3687},
numpages = {9},
location = {Virtual Event, Singapore},
series = {KDD '21}
}

@inproceedings{Guo2023com,
author = {Guo, Baoshen and Wang, Shuai and Wang, Haotian and Liu, Yunhuai and Kong, Fanshuo and Zhang, Desheng and He, Tian},
title = {Towards Equitable Assignment: Data-Driven Delivery Zone Partition at Last-mile Logistics},
year = {2023},
isbn = {9798400701030},
publisher = {Association for Computing Machinery},
address = {New York, NY, USA},
url = {https://doi.org/10.1145/3580305.3599915},
doi = {10.1145/3580305.3599915},
booktitle = {Proceedings of the 29th ACM SIGKDD Conference on Knowledge Discovery and Data Mining},
pages = {4078–4088},
numpages = {11},
location = {Long Beach, CA, USA},
series = {KDD '23}
}

@ARTICLE{Hao2024com,
  author={Miao, Hao and Zhong, Xiaolong and Liu, Jiaxin and Zhao, Yan and Zhao, Xiangyu and Qian, Weizhu and Zheng, Kai and Jensen, Christian S.},
  journal={IEEE Transactions on Knowledge and Data Engineering}, 
  title={Task Assignment With Efficient Federated Preference Learning in Spatial Crowdsourcing}, 
  year={2024},
  volume={36},
  number={4},
  pages={1800-1814},
  doi={10.1109/TKDE.2023.3311816}}

@ARTICLE{Qi2024com,
  author={Qi, Haoqing and Li, Yong and Xu, Yongqing and Quek, Tony Q. S.},
  journal={IEEE Wireless Communications Letters}, 
  title={UAV-Assisted Mobile Crowdsensing Systems Empowered by Wireless Power Transfer and Adaptive Compression Techniques}, 
  year={2024},
  volume={13},
  number={9},
  pages={2487-2491},
  doi={10.1109/LWC.2024.3421314}}

@inproceedings{Nayyar2025com,
author = {Nayyar, Rashmeet Kaur and Srivastava, Siddharth},
title = {Autonomous option invention for continual hierarchical reinforcement learning and planning},
year = {2025},
isbn = {978-1-57735-897-8},
publisher = {AAAI Press},
url = {https://doi.org/10.1609/aaai.v39i18.34163},
doi = {10.1609/aaai.v39i18.34163},
booktitle = {Proceedings of the Thirty-Ninth AAAI Conference on Artificial Intelligence and Thirty-Seventh Conference on Innovative Applications of Artificial Intelligence and Fifteenth Symposium on Educational Advances in Artificial Intelligence},
articleno = {2190},
numpages = {9},
series = {AAAI'25/IAAI'25/EAAI'25}
}

@INPROCEEDINGS{Habib2025com,
  author={Habib, Md Arafat and Iturria Rivera, Pedro Enrique and Ozcan, Yigit and Elsayed, Medhat and Bavand, Majid and Gaigalas, Raimundus and Erol-Kantarci, Melike},
  booktitle={2025 IEEE Wireless Communications and Networking Conference (WCNC)}, 
  title={LLM-Based Intent Processing and Network Optimization Using Attention-Based Hierarchical Reinforcement Learning}, 
  year={2025},
  volume={},
  number={},
  pages={1-6},
  doi={10.1109/WCNC61545.2025.10978505}}

@article{Ling2025com,
title = {Balancing supply and demand for ride-hailing: A preallocation hierarchical reinforcement learning approach},
journal = {Information Sciences},
volume = {718},
pages = {122371},
year = {2025},
issn = {0020-0255},
doi = {https://doi.org/10.1016/j.ins.2025.122371},
url = {https://www.sciencedirect.com/science/article/pii/S0020025525005031},
author = {Jiahao Ling and Xiaohui Huang and Xiaofei Yang and Boxue Cheng}
}

@article{Ruan2020base,
author = {Ruan, Sijie and Bao, Jie and Liang, Yuxuan and Li, Ruiyuan and He, Tianfu and Meng, Chuishi and Li, Yanhua and Wu, Yingcai and Zheng, Yu},
title = {Dynamic Public Resource Allocation Based on Human Mobility Prediction},
year = {2020},
issue_date = {March 2020},
publisher = {Association for Computing Machinery},
address = {New York, NY, USA},
volume = {4},
number = {1},
url = {https://doi.org/10.1145/3380986},
doi = {10.1145/3380986},
journal = {Proc. ACM Interact. Mob. Wearable Ubiquitous Technol.},
month = mar,
articleno = {25},
numpages = {22}
}

@article{lyu2026ts,
  title={TS-Memory: Plug-and-Play Memory for Time Series Foundation Models},
  author={Lyu, Sisuo and Zhong, Siru and Chen, Tiegang and Ruan, Weilin and Liu, Qingxiang and Lv, Taiqiang and Wen, Qingsong and Wong, Raymond Chi-Wing and Liang, Yuxuan},
  journal={arXiv preprint arXiv:2602.11550},
  year={2026}
}

@inproceedings{lyu2026occamvts,
  title={OccamVTS: Distilling Vision Models to 1\% Parameters for Time Series Forecasting},
  author={Lyu, Sisuo and Zhong, Siru and Ruan, Weilin and Liu, Qingxiang and Wen, Qingsong and Xiong, Hui and Liang, Yuxuan},
  booktitle={Proceedings of the AAAI Conference on Artificial Intelligence},
  volume={40},
  number={29},
  pages={24216--24225},
  year={2026}
}

\appendix

\section{Cooperative Air-Ground Delivery Platform}
\label{app:platform}

To bridge the gap between theoretical algorithms and real-world deployment, we developed a simulation platform. This platform is designed to evaluate algorithmic effectiveness using real-world urban road networks, such as Shanghai dataset~\cite{Ding2021dat} and allows for comprehensive benchmarking against various baselines, ranging from greedy approaches~\cite{Gao2024rel} to branch-and-bound strategies~\cite{Ruan2020base}. As illustrated in Figure~\ref{fig:agbench_structure1} and Figure~\ref{fig:agbench_structure2}, the system interface is composed of four modules: Scenario Configuration, Process Monitoring, Microscopic Agent Visualization, and Comparative Analytics.

\vspace{-1.0em}
\subsection{Scenario Configuration (Panel A)}
The \textbf{Configuration View} (Fig.~\ref{fig:agbench_structure1}-A) serves as the control center for initializing the research environment. It provides a modularized interface for customizing simulation parameters:
\begin{itemize}[leftmargin=*]
    \item \textbf{Environment Settings:} Researchers can define the global simulation context, including the road network dataset, order generation distribution, and the dispatching algorithm to be evaluated.
    \item \textbf{Heterogeneous Fleet Config:} The panel allows precise parameterization of the heterogeneous fleet. For UAVs, users can configure the fleet size, payload capacity, battery constraints, and maximum flight altitude. Similarly, for carriers, parameters such as speed limits, capacity, and service range can be adjusted.
\end{itemize}
This parametric design ensures that the platform can simulate diverse operational paradigms, from pure drone-based or rider-based models to complex collaborative air-ground scenarios.

\vspace{-1.0em}
\subsection{Delivery Process Monitoring (Panel B)}
\begin{figure}[t]
    \centering
\includegraphics[width=0.4\textwidth]{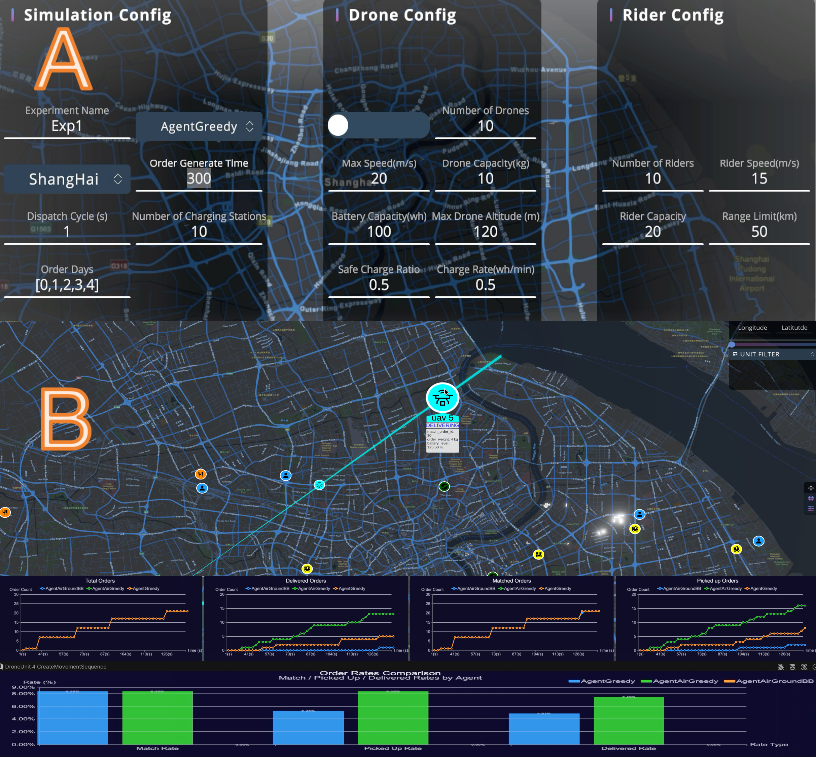}
    \caption{Configuration View and Delivery Process View.}
    \label{fig:agbench_structure1}
     \vspace{-1.0em}
\end{figure}

Upon initialization, the \textbf{Delivery Process View} (Fig.~\ref{fig:agbench_structure1}-B) visualizes the spatiotemporal dynamics of the entire system. 
The interface renders the real-time positions of all entities, including orders, UAVs, riders, and charging stations. 
The simulation unfolds in three sequential stages:
1) \textbf{Dispatching:} Orders are assigned to specific vehicles based on the selected algorithm;
2) \textbf{Pickup \& Delivery:} Vehicles execute route planning based on their specific mobility logic (Euclidean for UAVs vs. Road Network for riders) to complete tasks;
3) \textbf{Maintenance:} Crucially, the system incorporates a safety-aware logic for UAVs. After delivery, a UAV checks its residual energy. It will only accept new orders if the battery permits; otherwise, it autonomously navigates to the nearest charging station, simulating the operation constraints in real-world logistics.

\vspace{-1.0em}
\subsection{Microscopic Agent Visualization (Panel C)}
\begin{figure}[t]
    \centering
\includegraphics[width=0.4\textwidth]{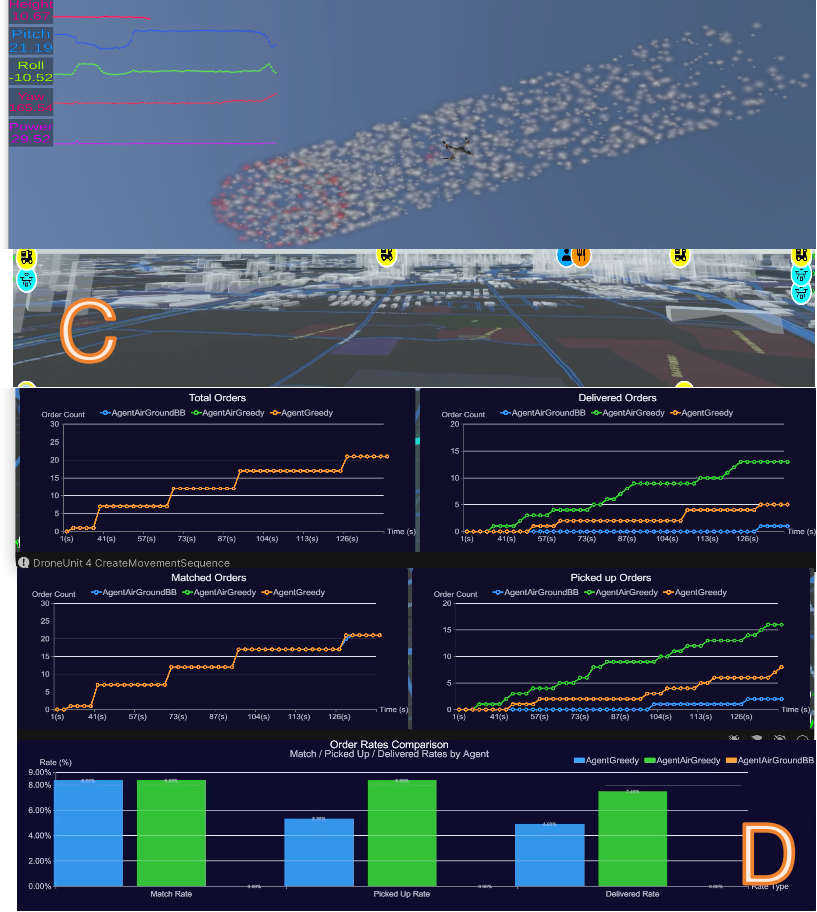}
    \caption{3D UAV View and Scheduling Result View.}
    \label{fig:agbench_structure2}
     \vspace{-1.0em}
\end{figure}

To support fine-grained analysis, the platform offers an interactive \textbf{Agent Perspective View} (Fig.~\ref{fig:agbench_structure2}-C). 
By clicking on individual entities (e.g., a specific UAV), the system switches to a microscopic telemetry view. This panel displays high-fidelity state information, including the agent's ID, current coordinates, and for UAVs, critical flight dynamics such as Height, Pitch, Roll, Yaw, and Remaining Power. 
This feature allows researchers to debug algorithmic behavior at the individual agent level and verify that physical constraints (e.g., battery depletion) are strictly observed.

\vspace{-1.0em}
\subsection{Real-time Comparative Analytics (Panel D)}
The \textbf{Result Dashboard} (Fig.~\ref{fig:agbench_structure2}-D) provides a quantitative assessment of algorithmic performance. 
It features a visualization system:
\begin{itemize}[leftmargin=*]
    \item \textbf{Dynamic Line Charts:} These tracks time-series metrics, plotting the number of generated orders against the volume of matched, picked-up, and delivered orders at each time step.
    \item \textbf{Summary Bar Charts:} These display aggregate performance indicators, such as overall matching rate and delivery rate.
\end{itemize}
A key feature of this module is its \textbf{time-synchronized comparison capability}. The system archives the performance traces of executed baselines. When a new algorithm is tested, its real-time results are overlaid with historical data, enabling researchers to visually benchmark the proposed method against state-of-the-art baselines in the same temporal context.










\end{document}